\documentclass[a4paper,fleqn,usenatbib]{mnras}

\usepackage{newtxtext,newtxmath}

\usepackage[T1]{fontenc}
\usepackage{ae,aecompl}

\usepackage{graphicx}	
\usepackage{amsmath}	
\usepackage{amssymb}	

\title[Gas accretion and the M-Z relation]
      {Gas accretion regulates the scatter of the mass-metallicity relation} 

\author[G.~De Lucia et al.]
       {Gabriella De Lucia$^{1}$\thanks{Email: gabriella.delucia@inaf.it},
        Lizhi Xie$^{2}$,
        Fabio Fontanot$^{1,3}$, 
        and Michaela Hirschmann$^4$\\
        $^1$INAF - Astronomical Observatory of Trieste, via G.B. Tiepolo 11, 
        I-34143 Trieste, Italy\\
        $^2$Tianjin Astrophysics Center, Tianjin Normal University,
        Binshuixidao 393, 300384, Tianjin, China\\
        $^3$IFPU  - Institute for Fundamental Physics of the Universe,
        via Beirut 2, 34151, Trieste, Italy\\
        $^4$DARK, Niels Bohr Institute, University of Copenhagen, Lyngbyvej 2,
        DK-2100 Copenhagen, Denmark}

\date{Accepted XXX. Received YYY; in original form ZZZ}

\pubyear{2020}

\begin{document}
\label{firstpage}
\pagerange{\pageref{firstpage}--\pageref{lastpage}}
\maketitle

\begin{abstract}
  In this paper, we take advantage of the GAlaxy Evolution and Assembly (GAEA)
  semi-analytic model to analyse the origin of secondary dependencies in the
  local galaxy mass - gas metallicity relation. Our model reproduces quite well
  the trends observed in the local Universe as a function of galaxy star
  formation rate and different gas-mass phases. We show that the cold gas
  content (whose largest fraction is represented by the atomic gas phase) can
  be considered as the third parameter governing the scatter of the predicted
  mass-metallicity relation, in agreement with the most recent observational
  measurements. The trends can be explained with fluctuations of the gas
  accretion rates: a decrease of the gas supply leads to an increase of the gas
  metallicity due to star formation, while an increase of the available cold
  gas leads to a metallicity depletion. We demonstrate that the former process
  is responsible for offsets above the mass-metallicity relation, while the
  latter is responsible for deviations below the mass-metallicity relation. In
  low and intermediate mass galaxies, these negative offsets are primarily
  determined by late gas cooling dominated by material that has been previously
  ejected due to stellar feedback. 
\end{abstract}


\section{Introduction}
\label{sec:intro}
Our quest to understand the formation and evolution of galaxies often proceeds
through the study of well defined correlations between their physical
properties. One of these relates the total galaxy stellar mass to the gaseous
metallicity. The former quantity represents the integrated mass of gas that has
been locked up in stars, while the latter reflects the recycling of gas and
metals from stars and the exchanges of gas between the galaxy and its
surroundings (inflows and outflows). First observations of a correlation
between gas metallicity and galaxy luminosity (as a proxy for galaxy mass) date
back to the 1970s (\citealt*{McClure_and_vandenBergh_1968};
\citealt{Lequeux_etal_1979}), and have been followed by more detailed studies
based on larger samples \citep*[e.g.][]{Skillman_etal_1989,Zaritsky_etal_1994}.
The advent of large spectroscopic surveys, and the development of more
sophisticated stellar population models have allowed major advances in our
ability to measure galaxy physical properties. The benchmark on the subject has
been provided by \citet{Tremonti_etal_2004}, who used $\sim 53,000$ galaxies
from the Sloan Digital Sky Survey (SDSS) to demonstrate the existence of a very
tight relation between the gas-phase metallicity and galaxy stellar mass. The
shape and normalization of the relation depend on the metallicity calibration
adopted \citep{Kewley_and_Ellison_2008}, which has to be taken into account
carefully when interpreting e.g. the evolution of the relation to higher
redshift or when comparing data from different studies. A tight correlation is
found up to $z\sim 3.5$, i.e. the highest redshift where all main optical lines
are still in the near-IR \citep[e.g.][just to cite a few]{Maiolino_etal_2008,Troncoso_etal_2014,Onodera_etal_2016,Sanders_etal_2020}.

Different explanations have been proposed for the origin of the
mass-metallicity relation. The classical interpretation is based on selective
losses of metals into the inter-galactic medium by low-mass galaxies, due to
galactic winds generated by the energy released by massive stars and supernovae
\citep[e.g.][]{Larson1974,Tremonti_etal_2004,Chisholm_etal_2018}. An
alternative possibility is that the observed relation reflects different
evolutionary stages of galaxies with different mass: high-mass galaxies evolve
more rapidly at higher redshift than their lower mass counterparts, that still
have to convert most of their gas into stars
\citep[e.g.][]{Maiolino_etal_2008,Zahid_etal_2011}. In addition, given the low
star formation rates measured in low-mass galaxies, infall of pristine gas is a
viable mechanism to dilute their interstellar medium 
\citep{Koeppen_and_Edmunds_1999,Dalcanton_2007}. Finally, variations of the
stellar initial mass function (IMF) could lead to systematic differences of the
metal yields as a function of the galaxy stellar mass
\citep*{Koeppen_etal_2007}. In reality, of course, all these effects are likely
simultaneously present, so that a correct interpretation of the observed trends
requires theoretical models that are fully embedded in a cosmological framework,
and that include all the relevant physical processes. 

The existence of a \emph{secondary dependence} in the mass-metallicity relation
was already studied in \citet{Tremonti_etal_2004}. These authors found that the
residuals from the best fit relation correlate with the local surface mass
density measured within the fibre aperture and somewhat with colours, but not
with the H$_\alpha$ equivalent width (see their Fig.~7). Later studies, based
on the same data sample \citep{Mannucci_etal_2010,Lara-Lopez_etal_2010}, have
identified a strong dependence on the star formation rate, that has been framed
in terms of a more general relation between galaxy stellar
mass, gas-phase metallicity, and star formation rate. This three-dimensional
relation can be explained in terms of a more fundamental relation with the gas
content of galaxies (or gas fraction): larger amounts of gas lead to increasing
star formation rates and therefore to increasing values of the gaseous
metallicity, while a decreasing metallicity is expected with dilution caused by
larger amounts of newly accreted gas
\citep[e.g.][]{Dayal_etal_2013,Lilly_etal_2013}. A strong correlation between
gas fraction/gas content and metallicity, at fixed stellar mass, has indeed
been observed in the local Universe
\citep{Hughes_etal_2013,Bothwell_etal_2016,Brown_etal_2018}. These studies have
shown that the correlation with gas content is significantly stronger than that
with the star formation rate (either total or measured within a fibre),
independently of the metallicity calibration adopted. While outflows can
contribute to the scatter in the observed mass-metallicity relation, the
observed trends suggest that it is primarily driven by fluctuations in
accretion rates of (pristine) gas (see discussion in \citealt{Brown_etal_2018}).
This picture is supported by models assuming `equilibrium' solutions in which
gas inflow is compensated by star formation and outflows
\citep[e.g.][]{Lilly_etal_2013,Dave_etal_2012}. This formalism is useful to
understand the origin of the main trends and appears to be able to broadly
capture, at least to first order, results from more complex theoretical models
of galaxy formation. It is, however, not appropriate to characterize the
contribution from different physical processes to the scatter in the observed
scaling relations, and to fully account for the variety of galaxy evolutionary
histories driven by structure formation. 

Cosmological models of galaxy formation (both semi-analytic models and
hydro-dynamical simulations) have long had problems in reproducing well the
shape of the observed mass-metallicity relation at $z=0$ and, in particular,
its evolution to higher redshift (see discussion in
\citealt{Somerville_and_Dave_2015}, and their Fig.~6). More recent renditions
of these theoretical models exhibit a significantly better agreement with data
\citep{Hirschmann_etal_2016,Xie_etal_2017,DeRossi_etal_2017,Collacchioni_etal_2018,Torrey_etal_2019}. In some cases, an analysis of the scatter of the
mass-metallicity relation and its origin has been carried out.
\citet{Yates_etal_2012} studied the dependence of the cold gas metallicity on
star formation rate, in an earlier version of the semi-analytic model
{\sc L-Galaxies} \cite[in particular, they use the model published
  in][]{Guo_etal_2011}. They find a star formation rate dependence that is
qualitatively similar to that observed, but their work focuses on the reversal
of the dependence for low and high-mass galaxies. More recently,
\citet{Lagos_etal_2016} have used a principal component analysis to show that
the atomic gas fraction, stellar mass and star formation rate account for most
of the variance in the galaxy population predicted by the EAGLE simulations.
They argue that this `fundamental plane' arises from the self-regulation of the
star formation in galaxies. \citet{Torrey_etal_2019} analyse the evolution of
the mass-metallicity relation within the IllustrisTNG simulation suite. They
find a correlation between the scatter in the predicted relation and the
galactic gas-mass, that they argue can be explained as a competition between
periods of gas-rich, enrichment domination and periods of gas-poor, accretion
domination. To date, no systematic comparison has been carried out between
predictions of the most recent theoretical models and the accurate measurements
available in the local Universe for trends as a function
of {\it both} star formation rates and the different gas phase components. In
addition, a detailed quantification of the physical processes responsible for
the predicted trends has not been published yet.

In this paper, we take advantage of a state-of-the-art semi-analytic model that
includes an accurate treatment for the non-instantaneous recycling of metals,
gas and energy, as well as a treatment for the partition of the cold gas in
atomic and molecular hydrogen, and analyse the origin of secondary dependencies
in the mass-metallicity relation at $z=0$. A brief overview of the model used
in this work is given in Section~\ref{sec:sam}. In Section~\ref{sec:mzsec}, we
present the basic predictions of our model and compare them with recent
observational measurements. In Section~\ref{sec:mz}, we analyse the
contribution of different physical processes to the scatter of the
mass-metallicity relation. Finally, in Section~\ref{sec:discconcl}, we discuss
our results and give our conclusions. 


\section{The galaxy formation model}
\label{sec:sam}

In this work, we use an updated version of the GAlaxy Evolution and Assembly
({\sc GAEA}) semi-analytic model \citep{Hirschmann_etal_2016}. {\sc GAEA}
descends from the model originally published in
\citet{DeLucia_and_Blaizot_2007}, but the treatment of various physical
processes has been significantly updated since then. In particular, the {\sc
  GAEA} version used in this study includes (i) a sophisticated scheme to
account for the non-instantaneous recycling of gas, energy and metals
\citep{DeLucia_etal_2014}, that allows us to trace the evolution of individual
metal abundances; (ii) a stellar feedback scheme that is partly based on
hydrodynamical simulations, and that allows us to reproduce the evolution of
the galaxy stellar mass function up to $z\sim 7$ and the cosmic star formation
rate density up to $z\sim 10$ \citep{Hirschmann_etal_2016,Fontanot_etal_2017};
and (iii) a self-consistent treatment to partition the cold gas in its atomic
and molecular hydrogen components \citep{Xie_etal_2017}, that will allow us to
analyse explicitly the dependence of the mass-metallicity relation on different
components of the cold gaseous phase. Specifically, we use here the
parametrizations based on the empirical prescriptions by
\citet{Blitz_and_Rosolowsky_2006} that, in the framework of our model, provide
a good agreement with the scaling relations observed in the local Universe. We
refer to the original papers mentioned above for full details on the modelling
adopted for different physical processes. Briefly, we assume that infalling gas
has primordial chemical composition and condenses at the centre of dark matter
haloes, settling in a gaseous disk whose radius is estimated by tracing the
angular momentum evolution of the gas \citep{Xie_etal_2017}. In the regions of
highest gas densities, atomic gas can be efficiently converted in molecular
hydrogen that represents, in our model, the fuel for star formation. The energy
from supernovae explosions and massive stars drives efficient galactic scale
winds that can eject significant amounts of gas and metals outside the parent
dark matter haloes. The ejected gas can later be re-incorporated, on
time-scales that are assumed to depend on halo mass \citep[][and references
  therein]{Hirschmann_etal_2016}. In our model, we always assume that metal
flows occur proportionally to mass flows between different galactic components.

In previous work, we have shown that our reference {\sc GAEA} model is able to
reproduce a number of important observational constraints. Notably for the
present study, it is one of the few recently published theoretical models able
to reproduce the observed correlation between galaxy stellar mass and gas
metallicity, as well as its evolution as a function of redshift
\citep{Hirschmann_etal_2016,Xie_etal_2017}, without assuming the chemical yield
as a free parameter.  It therefore represents an ideal tool to investigate the
origin of secondary dependencies in such a correlation.

The model predictions presented in the following are based on merger trees
extracted from the Millennium Simulation \citep{Springel_etal_2005}. The
simulation follows 2,160$^3$ dark matter particles in a box of 500~${\rm
  Mpc}\,{\rm h}^{-1}$ on a side, and assumes cosmological parameters consistent
with WMAP1 ($\Omega_\Lambda=0.75$, $\Omega_m=0.25$, $\Omega_b=0.045$, $n=1$,
$\sigma_8=0.9$, and $H_0=73 \, {\rm km\,s^{-1}\,Mpc^{-1}}$). Recent
measurements provide slightly different cosmological parameters (in particular,
a larger value for $\Omega_m$ and a lower value for $\sigma_8$).  As we have
shown in previous work, however, these differences do not affect significantly
model predictions once the parameters of the model have been retuned to
reproduce a given set of observational measurements in the local Universe
\citep{Wang_etal_2008}. Simulation data were stored in 64 outputs, that are
approximately logarithmically spaced in time between $z=20$ and 1, and linearly
spaced in time for $z < 1$. The differential equations governing the evolution
of model galaxies are solved dividing the time interval between two subsequent
snapshots in $20$ equal sub-steps that, in our reference runs, correspond to a
maximum of $\sim 18.8$~Myr.

For the analysis presented below, we have used only about 6 per cent of the
volume of the simulation, but we have verified that results do not vary
significantly when considering a larger volume. The comparison with data
presented in Section~\ref{sec:mzsec} is based on 20 per cent of the volume of
the simulation.

\section{Secondary correlations in the (gas) metallicity - galaxy mass relation}
\label{sec:mzsec}

\begin{figure}
\centering
\resizebox{8.7cm}{!}{\includegraphics{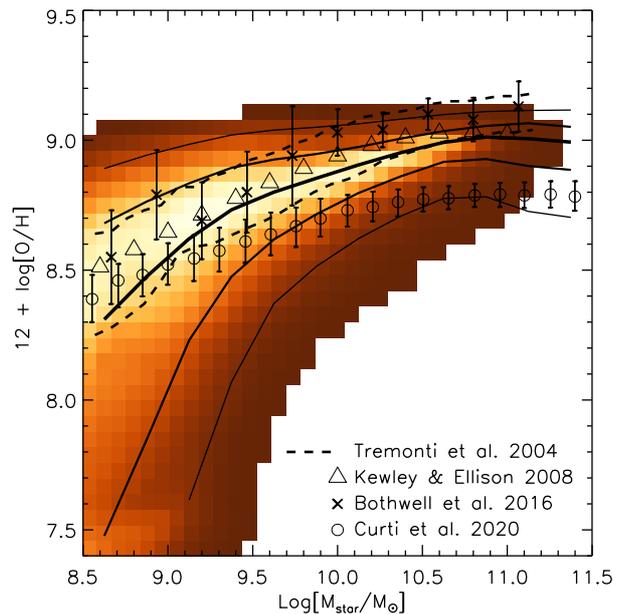}} 
\caption{Cold gas metallicity as a function of galaxy stellar mass, as
  predicted by our model (coloured distribution) and compared with different
  observational measurements (symbols and dashed lines, as indicated in the
  legend). The solid black lines correspond to the median and scatter
  (1-$\sigma$, and 2-$\sigma$ - ordered by decreasing thickness) of the
  predicted distribution.
\label{fig:mz}}
\end{figure}

Fig.~\ref{fig:mz} shows the gas metallicity-stellar mass relation predicted by
our model (coloured distribution and solid black lines), compared with
different observational measurements. In particular, dashed lines show the 16th
and 84th percentiles of the distribution measured by \cite{Tremonti_etal_2004}
using SDSS spectra from DR2, and fitting simultaneously all most prominent
emission lines. The empty triangles show a re-determination of the
mass-metallicity relation by \citet{Kewley_and_Ellison_2008}, based on spectra
from DR4 with tighter limits on the redshift range ($0.04 < z < 0.1$). With
this selection, the projected SDSS fibre covers more than 20 per cent of the
total g-band light, and incompleteness at higher redshift is minimized. The
empty triangles correspond to a metallicity calibration based on
photoionization models \citep{Kewley_and_Dopita_2002}, that suffers of similar
biases and uncertainties of the strong line method used by \citet[][see
  discussion in \citealt{Kewley_and_Ellison_2008} for
  details]{Tremonti_etal_2004}. Crosses with error bars show measurements by
\citet{Bothwell_etal_2016}, based on a sample of local galaxies with available
measurements of both molecular and total gas mass. Specifically, the data come
mainly from three surveys: COLD GASS \citep{Saintonge_etal_2011}, the Herschel
Reference Survey \citep{Boselli_etal_2014}, and ALLSMOG
\citep{Bothwell_etal_2014}. Also for this sample, metallicity estimates are
based on optical strong-line fluxes. Finally, open circles with error bars show
measurements by \citet{Curti_etal_2020}.  These authors have re-analysed the
SDSS data using a new metallicity calibration based on the T$_e$ abundance
scale \citep{Curti_etal_2017}. All observational measurements shown assume a
\citet{Kroupa_2001} stellar IMF, that is very similar to the
\citet{Chabrier_2003} IMF adopted in our reference model (in the case of Curti
et al., the stellar mass measurements have been already rescaled to a Chabrier
IMF). As mentioned in Section~\ref{sec:intro}, the shape and normalization of
the mass-metallicity relation depend significantly on the calibration
adopted. It is well known, and this is clearly shown in Fig.~\ref{fig:mz}, that
there are large systematic offsets between estimates based on theoretical
calibrations and electron temperature metallicities. The origin of these
offsets remains unclear, but it is generally accepted that calibrations based on
strong emission lines and photoionization models tend to over-estimate the true
metallicities, while methods based on the electron temperature can
under-estimate them \citep[e.g.][and references
  therein]{Kewley_and_Ellison_2008, Maiolino_and_Mannucci_2019}.

The solid black lines in Fig.~\ref{fig:mz} show the median, 1-$\sigma$, and
2-$\sigma$ scatter of the distribution predicted for model star forming
galaxies. This particular selection has been used to reflect the fact that our
main comparison samples are composed primarily of star-forming galaxies.
Specifically, we have selected model star forming galaxies imposing a cold gas
fraction of at least 25 per cent, and a specific star formation rate larger than
0.3/t$_{\rm H}$ (with t$_{\rm H}$ equal to the Hubble time). These galaxies
will represent the model sample used throughout this paper (unless otherwise
specified). We have verified that results do not change significantly by
increasing the threshold adopted for the gas fraction, but this tends to remove
the most massive galaxies, and preferentially those with largest metallicity
(as we will show below). As in \citet{Xie_etal_2017}, we remove Helium (26 per
cent of the total gas mass) to get the abundance of atomic hydrogen for our
model galaxies. The figure shows that model predictions are in quite good
agreement with observational measurements based on strong emission lines or
photoionization models, both in terms of normalization and overall shape of the
mass-metallicity relation. The predicted scatter is larger than observed,
particularly for galaxies with stellar mass smaller than $\sim 10^{10}\,{\rm
  M}_{\sun}$ and with metallicity below the median, i.e. the distribution of
model metallicities is somewhat more skewed than observed. 

\begin{figure*}
\centering
\resizebox{7.5cm}{!}{\includegraphics{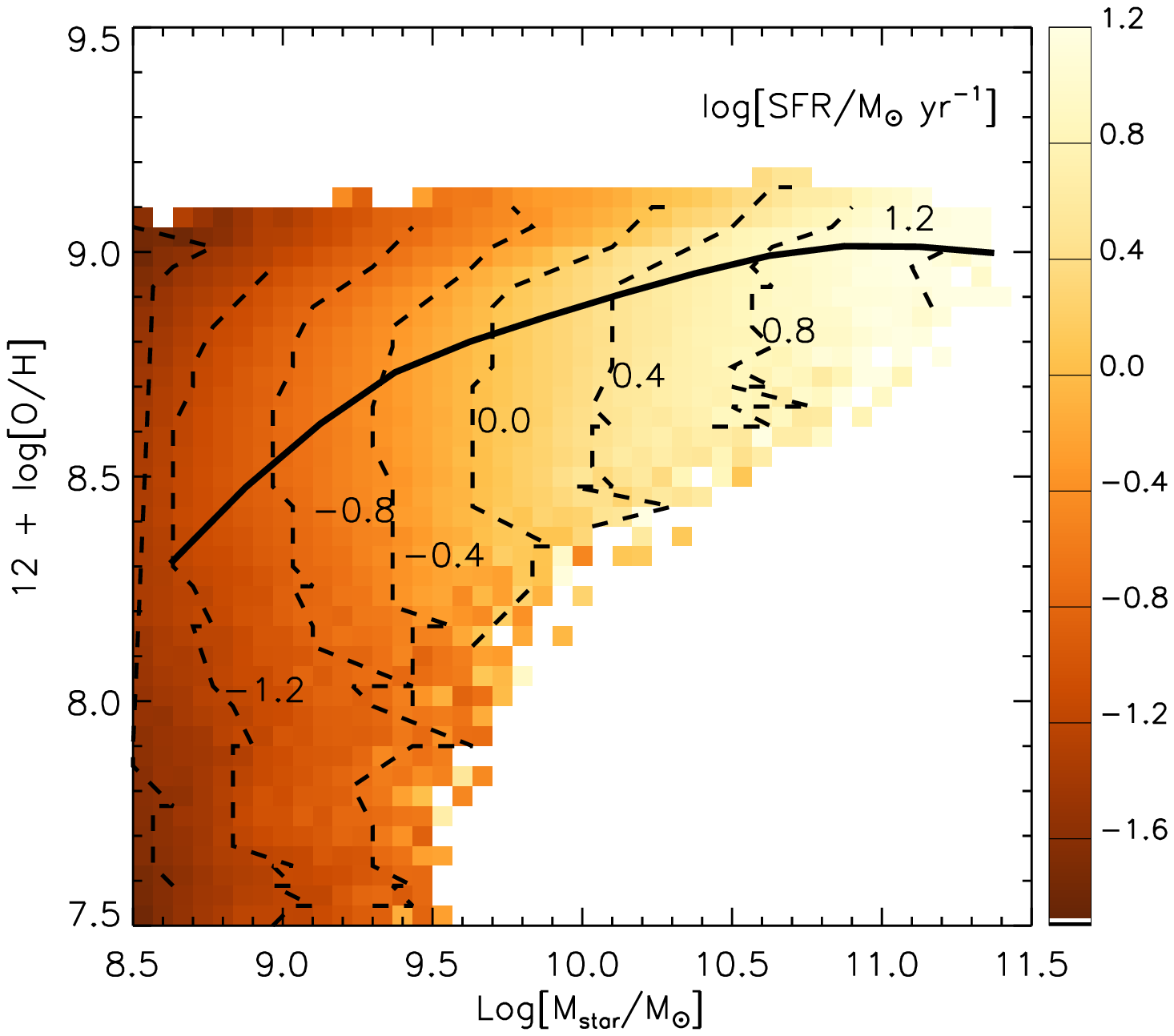}}
\hspace{1cm}
\resizebox{7.5cm}{!}{\includegraphics{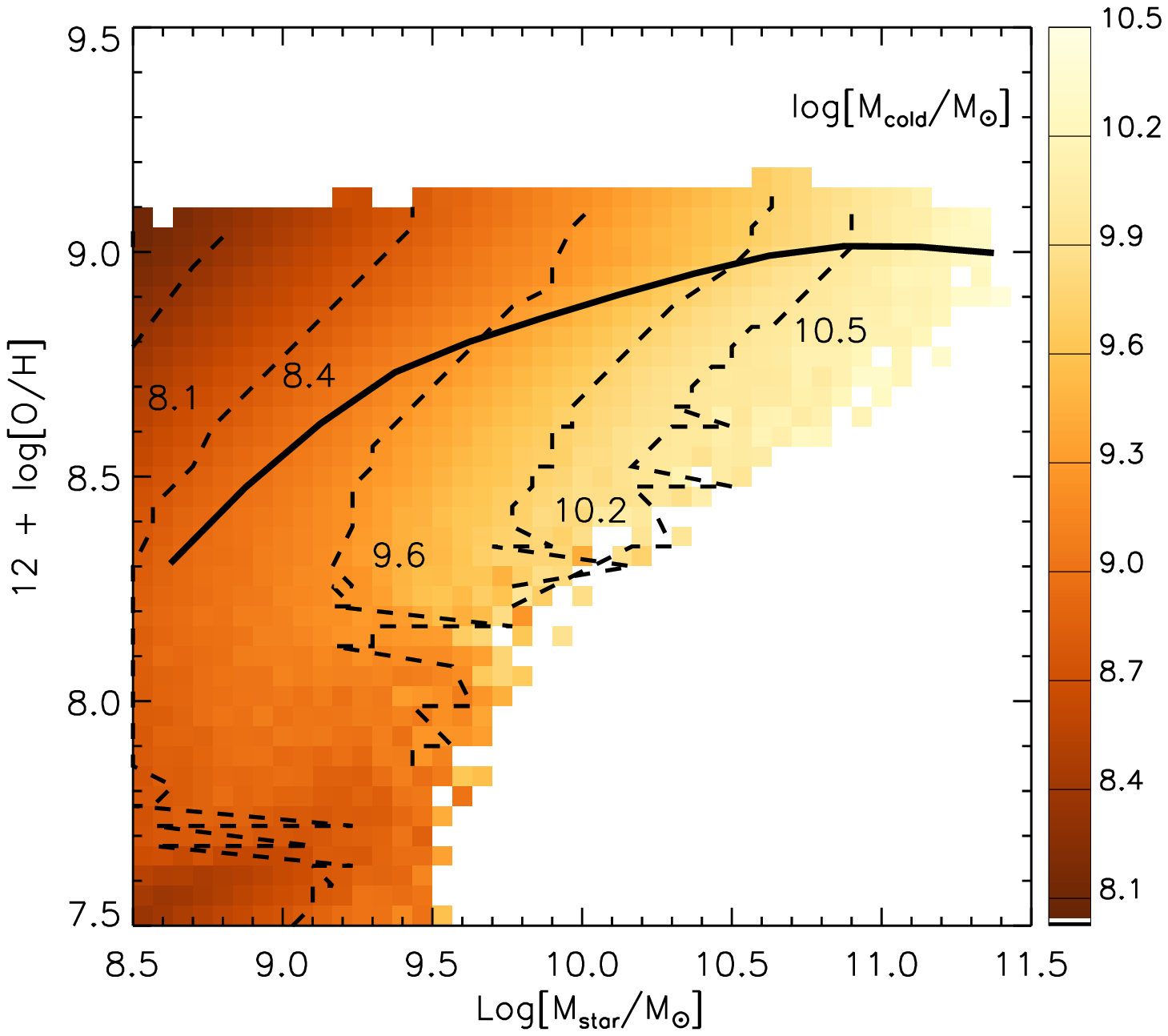}}
\resizebox{7.5cm}{!}{\includegraphics{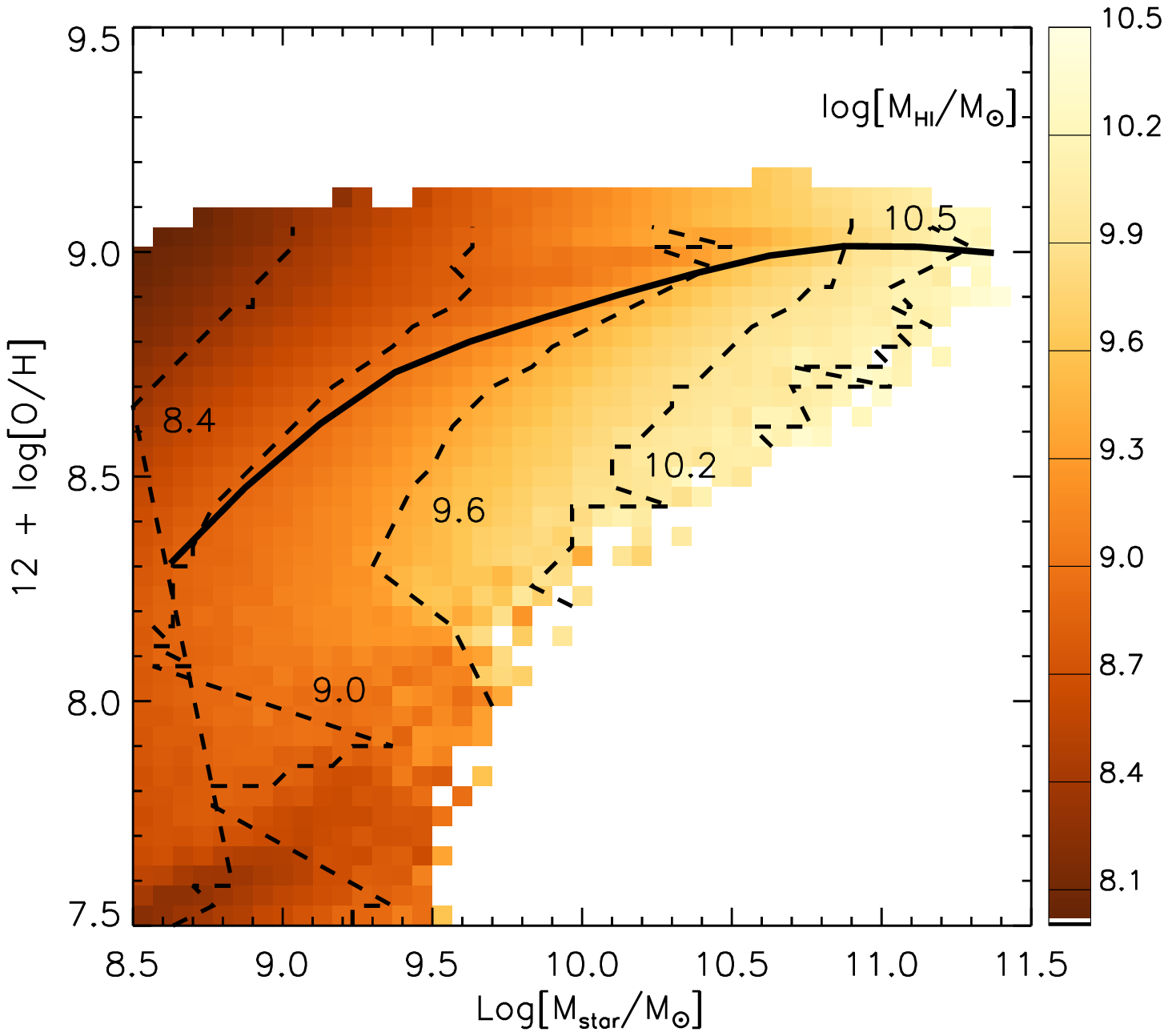}} 
\hspace{1cm}
\resizebox{7.5cm}{!}{\includegraphics{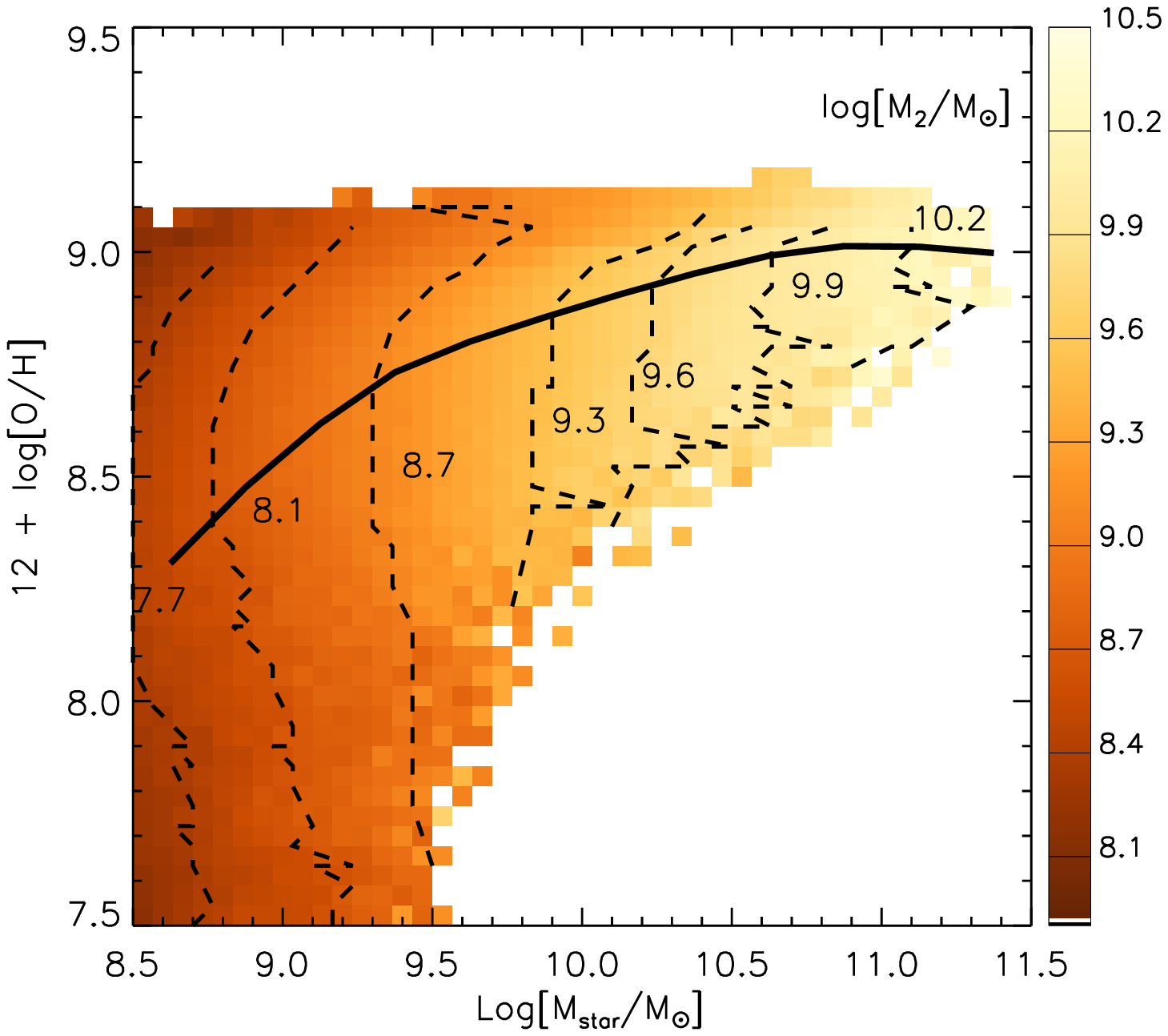}}
\caption{As in Fig.~\ref{fig:mz}, but colour-coded as a function of the median
  of the quantity indicated on the top right of each panel. The black solid
  line shows the median mass-metallicity relation obtained for all star-forming
  model galaxies considered (see text for details). Dashed lines show the
  location of star-forming galaxies whose median value of the quantity
  indicated on the top-right of the panel is close (within 0.3 dex) to that
  indicated close to each line.
  \label{fig:dep}}
\end{figure*}

\begin{figure}
\centering
\resizebox{8.7cm}{!}{\includegraphics{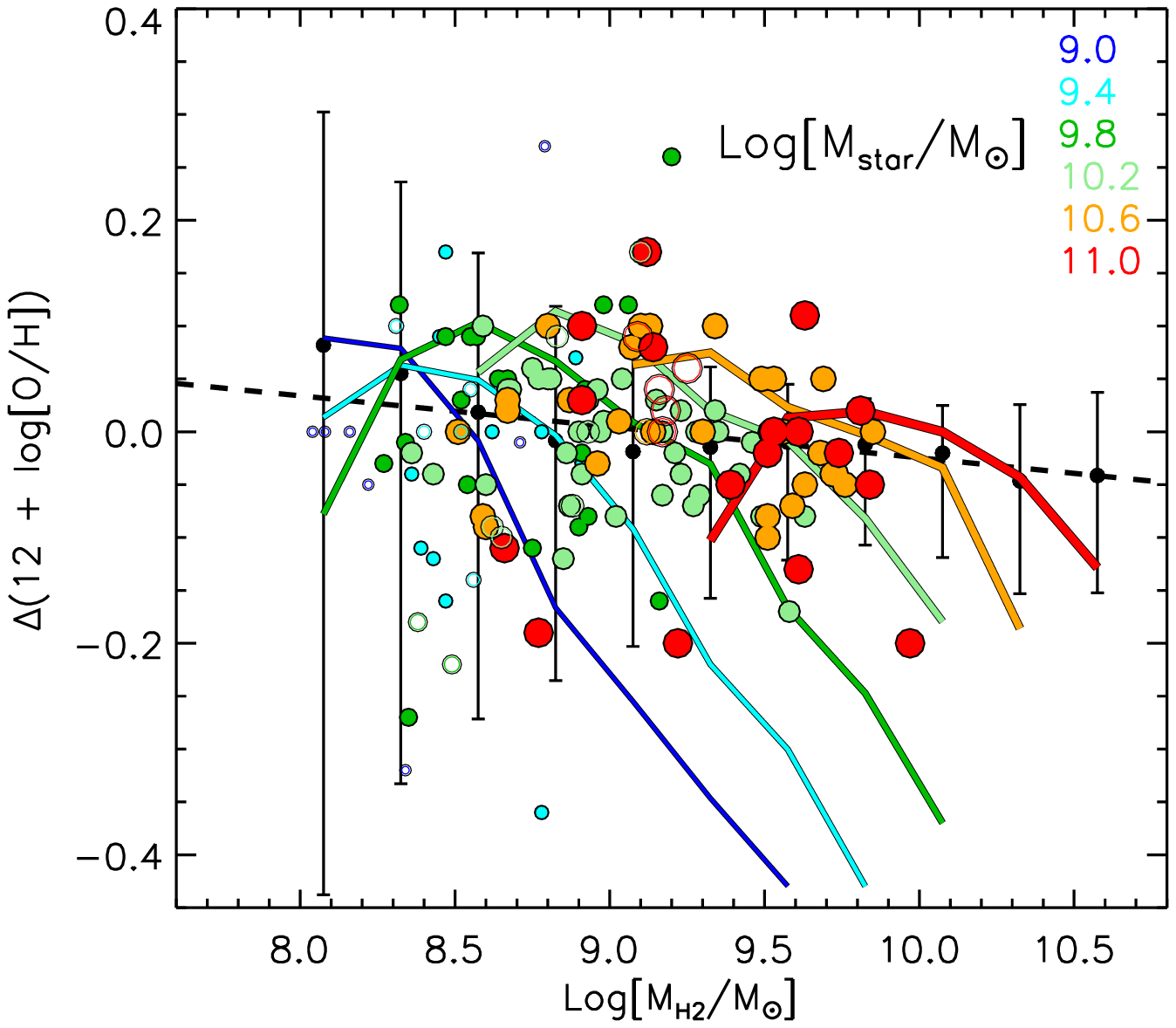}} 
\caption{Offset from the mass-metallicity relation as a function of the
  molecular hydrogen mass. Symbols show observational measurements by
  \citet{Bothwell_etal_2016}, and are colour-coded as a function of the galaxy
  stellar mass as indicated in the legend (the size of the symbols increases
  with galaxy stellar mass). Open symbols highlight data for which only upper
  limits for the molecular mass estimates are available. Solid lines, of
  thickness increasing with increasing galaxy stellar mass, show model
  predictions for the same galaxy mass bins considered for the data.  Black
  symbols with error bars show the median and percentiles (16th and 84th)
  obtained for the model sample, independently of the galaxy stellar mass. The
  dashed line represents a linear fit to the black circles.
\label{fig:both}}
\end{figure}

Fig.~\ref{fig:dep} shows the distribution of the same star forming model
galaxies in the galaxy stellar mass - gas metallicity plane but colour coded as
a function of the median value of the SFR (top left panel), cold gas mass (top
right panel), molecular gas mass (bottom left panel), and atomic gas mass
(bottom right panel). The figure clearly shows that all quantities considered
contribute to some extent to the scatter of the mass-metallicity
relation. Lines of constant star formation rate (and molecular gas content) are
almost vertical, i.e. at fixed galaxy stellar mass, there is only a weak
dependence of the cold gas metallicity on the star formation rate or on the
molecular gas content. The dependence is stronger for the total amount of gas
or the mass of atomic gas, that represents its largest fraction. It is not
surprising that the trends predicted for the star formation rate are very
similar to those obtained for the molecular gas as, in our model, these two
quantities are directly related \citep{Xie_etal_2017}. In the following, we
compare in more detail our model predictions with results from recent
observational work. 

Fig.~\ref{fig:both} shows the offset from the best-fit mass-metallicity
relation as a function of the molecular hydrogen mass. Observational
measurements by \citet{Bothwell_etal_2016} are shown as symbols and are
colour-coded as a function of galaxy stellar mass, as indicated in the legend.
Open symbols highlight data with only upper limits for the molecular mass
estimates. The latter are based on a CO to H$_2$ conversion factor that depends
on metallicity. Solid lines of different colours show model predictions for
the same galaxy stellar mass bins considered for the data. For both the model
and observational samples, metallicity offsets have been computed evaluating,
for each galaxy, the median metallicity of all galaxies in the sample with very
similar galaxy mass (within $\pm 0.05$ dex) and the corresponding difference
from the predicted and observed best-fit mass-metallicity relation,
respectively. As in Fig.~\ref{fig:mz}, and for consistency with our model
sample, we have considered only galaxies in the local Universe from the sample
presented in \citet{Bothwell_etal_2016}. This leaves only about 160 galaxies,
which is significantly smaller than the number of model galaxies considered.
Albeit statistically small, the observed sample exhibits a weak trend of
increasing (negative) offsets with increasing gas mass, within each galaxy mass
bin. The trend is qualitatively similar to that obtained for model galaxies.
The dispersion of the metallicity offsets decreases with increasing molecular
mass, both for the observed and model galaxies. In addition, consistently with
findings by \citet{Bothwell_etal_2016}, we find a mild decrease of the average
metallicity offset with increasing molecular mass.

\begin{figure*}
\centering
\resizebox{16cm}{!}{\includegraphics{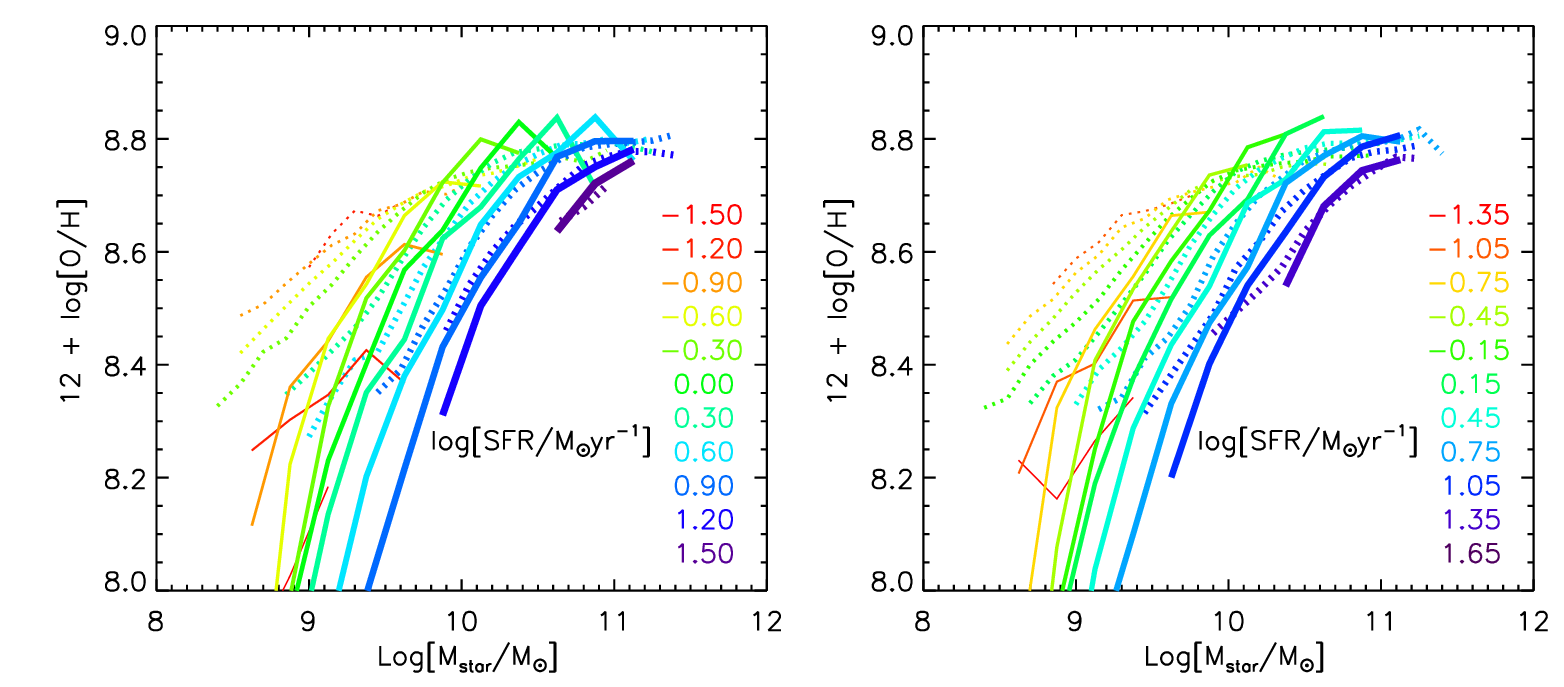}} 
\caption{Cold gas metallicity as a function of galaxy stellar mass, for bins of
  galaxies with different star formation rates as indicated in the legend.
  Dotted lines show observational measurements by \citet{Curti_etal_2020},
  while solid lines show predictions based on our reference model. The line
  thickness increases with increasing values of the star formation rate. A
  constant vertical shift of $-0.22$ has been applied to all model
  predictions. These have {\it not} been convolved with observational
  uncertainties. When this is done, the dependence on star formation rates
  disappears for the 5-6 bins with lowest star formation rates, while the
  trends shown for bins with larger star formation rates are not significantly
  affected.
\label{fig:curtisfr}}
\end{figure*}

\begin{figure*}
\centering
\resizebox{16cm}{!}{\includegraphics{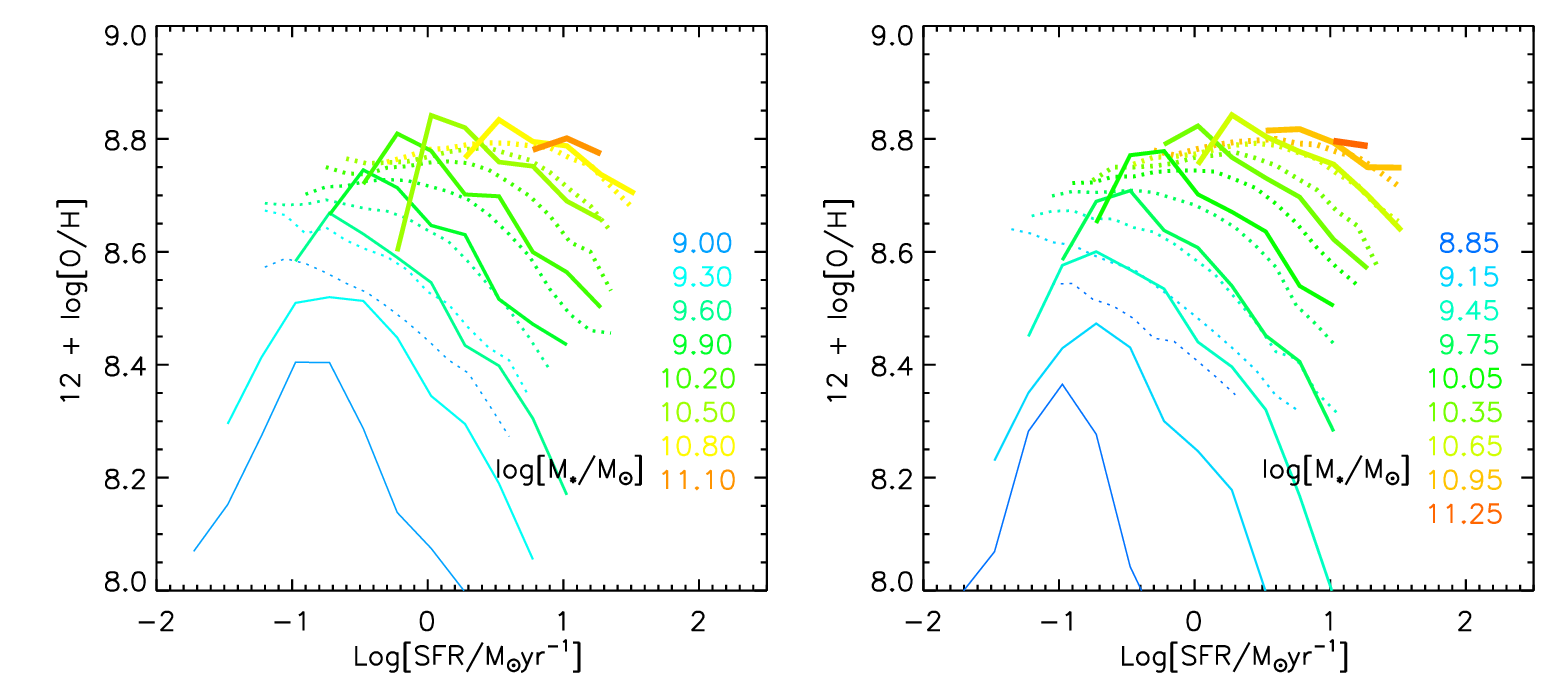}} 
\caption{Cold gas metallicity as a function of galaxy star formation rate, for
  bins of galaxy stellar mass as indicated in the legend. Dotted lines show
  observational measurements by \citet{Curti_etal_2020}, while model
  predictions are shown as solid lines. The line thickness increases with
  increasing galaxy stellar mass. As in Fig.~\ref{fig:curtisfr}, a constant
  vertical shift of $-0.22$ has been applied to model predictions. These have
  {\it not} been convolved with observational uncertainties. We have verified
  that accounting for such uncertainties would flatten the trends at low star
  formation rates.
\label{fig:curtimstar}}
\end{figure*}

Figs.~\ref{fig:curtisfr} and \ref{fig:curtimstar} show a comparison with the
recent work by \citet{Curti_etal_2020}. In Fig.~\ref{fig:curtisfr}, both the
observed (dotted lines) and model galaxy samples (solid lines) are binned as a
function of the total galaxy star formation rate, as indicated by the legend. A
constant vertical shift of $-0.22$ has been applied to model predictions
(the value has been chosen to reproduce approximately the normalization of the
relation observed for the most massive galaxies). In the data, the star
formation rate is determined from the extinction corrected H$\alpha$ luminosity
inside the SDSS fibre, and using aperture corrections by
\citet{Salim_etal_2007}. Observational estimates for both stellar masses and
star formation rates have been re-scaled to a Chabrier IMF, that is the same
adopted in our model. The figure shows that model predictions follow
qualitatively the observed trends, but the significance of the SFR dependence
is somewhat weaker in the model, that also predicts a steeper slope of the
mass-metallicity relation. The overall normalization is different: as discussed
above, predictions from our model are in better agreement with measurements
based on strong emission lines and photoionization methods. We have verified
that, accounting for typical observational uncertainties in the
measurements\footnote{In this case, we have simply convolved model predictions
  assuming a Gaussian uncertainty on ${\rm log\,M_{\star}}$ and ${\rm log\,SFR}$
  of 0.15-0.20.} of galaxy stellar masses and galaxy star formation rates,
removes the dependence predicted for the 5-6 bins with lowest star formation
rates (which is similar to the trends observed in the data).

\begin{figure*}
\centering
\resizebox{16cm}{!}{\includegraphics{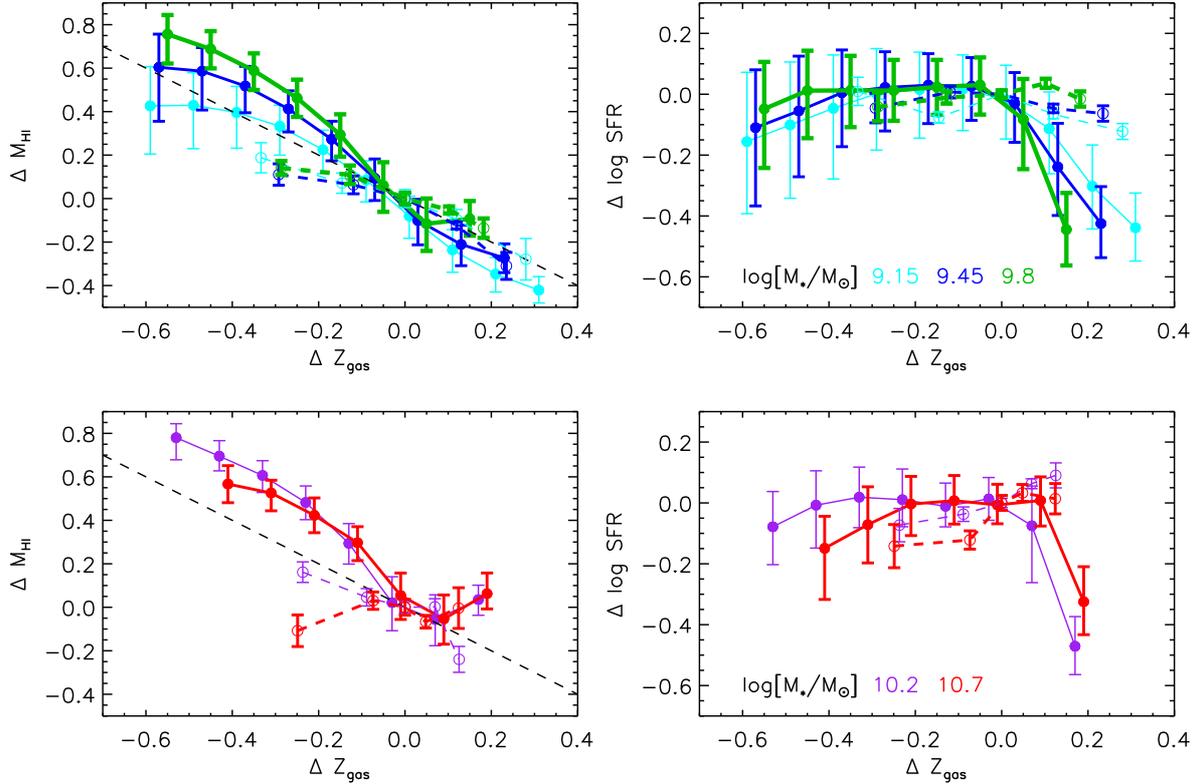}} 
\caption{Differences of HI mass (left panels) and star formation rate (right
  panels) as a function of the differences in metallicity from the value of the
  same quantities on the mass-metallicity relation. Top and bottom panels show
  different stellar mass bins, with model predictions shown as filled symbols
  connected by solid lines and observational estimates by
  \citep{Brown_etal_2018} marked by open circles connected by dashed lines.  In
  all panels, the line thickness increases with increasing galaxy stellar mass.
\label{fig:brown2}}
\end{figure*}

Fig.~\ref{fig:curtimstar} shows the relation between gaseous metallicity and
star formation rate, for different stellar mass bins. As in the previous
figure, observational measurements are shown as dotted lines and model
predictions as solid lines. Also in this case, we have applied a constant
vertical shift of $-0.22$. This works relatively well for the most massive bins
considered, but brings model predictions significantly below the observational
estimates for the least massive galaxies considered. Model predictions exhibit
an inversion at star formation rates below
${\rm Log(SFR/{\rm M_{\sun}\,yr^{-1}})}\sim -1$. Very few such low estimates are
available in the observational sample, and we have verified that the trends
would flatten when accounting for observational uncertainties on the estimate
of the star formation rate.

Fig.~\ref{fig:brown2} shows a comparison between our model predictions and
observational estimates by \citet{Brown_etal_2018} in terms of the differences
of a given parameter (in this specific case, the atomic hydrogen mass
M$_{\rm HI}$ and the star formation rate) from the value of the same quantity on
the mass-metallicity relation, at fixed galaxy stellar mass. Different stellar
mass bins are shown in the top and bottom panels, with observational estimates
marked by empty symbols connected by dashed lines and model predictions shown
as filled symbols connected by solid lines. The latter have been estimated with
respect to the best fit mass-metallicity relation measured for model galaxies.
We only show observational estimates corresponding to the total star formation
estimates by \citet{Salim_etal_2016} and to the metallicity calibration by
\citet{Tremonti_etal_2004}, that provides a good agreement with our predicted
mass-metallicity relation. In the left panels, the dashed lines correspond to
the inverse one-to-one relation, to guide the eye. Our model predicts trends
that are qualitatively similar to those observed, with a few important
differences, that can be summarized as follows:
\begin{itemize}
\item[(i)] the metallicity offsets predicted by our model extend to more
  negative values than the observational data, i.e. there are more significant
  deviations below the mass-metallicity relation in the model. This is
  consistent with the large scatter towards low metallicities shown for our
  model galaxies in Fig.~\ref{fig:mz};
\item[(ii)] for galaxies that are offset above the mass-metallicity relation
  (positive values of $\Delta\,Z_{\rm gas}$), our model predicts a stronger
  dependence on the star formation rate than observed, with smaller (more
  negative) offsets from the average values of the star formation rate on the
  mass-metallicity relation.
\end{itemize}

\section{Which physical processes drive the observed secondary dependencies?}
\label{sec:mz}

\begin{figure*}
\centering
\resizebox{16cm}{!}{\includegraphics{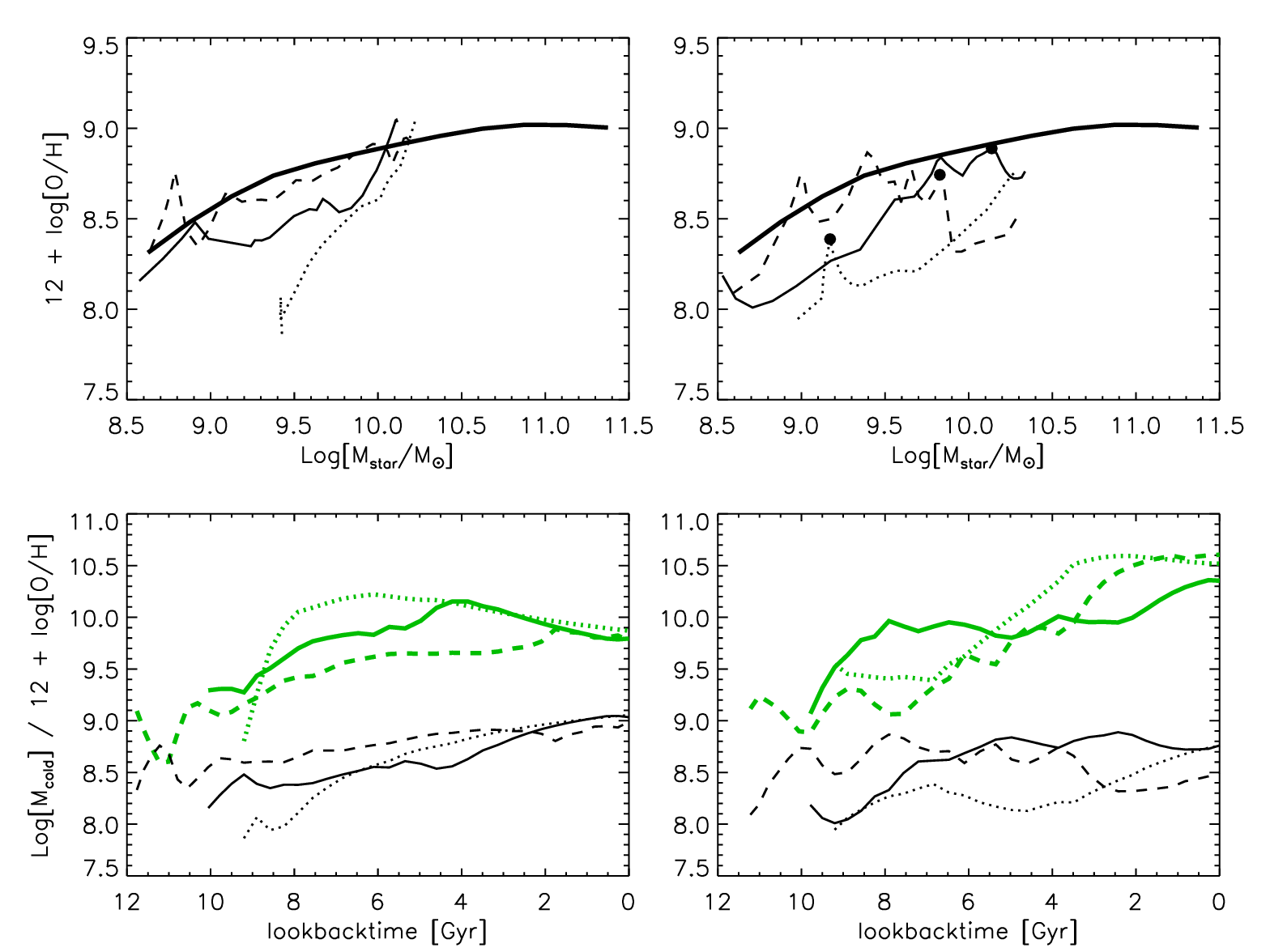}} 
\caption{Top panels show the trajectories of a few example galaxies in the
  stellar mass - gaseous metallicity plane. Left and right panels are for
  galaxies offset above and below the mass-metallicity relation, respectively.
  Dashed and dotted lines correspond to example galaxies whose trajectories are
  relatively close or distant from the predicted relation at z=0 (thick solid
  line) for a significant fraction of the galaxy's lifetime. Solid lines
  correspond to intermediate cases (the most frequent among the model
    galaxies analysed - see text). The bottom panels show the corresponding
  evolution of the cold gas mass (thick green lines) and gaseous metallicity
  (black lines). The filled circles in the top right panel mark the first
  metallicity maximum, going backwards in time, identified for each model
  galaxy (this will be used later in our analysis).
\label{fig:exhist}}
\end{figure*}

The results discussed above show that our model provides a good agreement with
observational measurements of the mass-metallicity relation in the local
Universe, as well as of the related secondary dependence trends. The agreement
is not perfect, but the main trends are qualitatively reproduced. This confirms
that our model represents an ideal tool to investigate their origin in a
cosmological context, and to quantify the relative importance of different
physical processes. 

In order to characterize the evolution of our model galaxies and identify the
physical processes that are driving the trends discussed above, we have
analysed in detail the evolution of individual model galaxies by following
their main progenitor branch, and saving all relevant information about the
flows of gas and metals driven by different physical processes. For the
analysis presented in this section, we have focused only on central
galaxies. These represent, overall, the largest fraction ($\sim 67$ per cent) of
the model star forming galaxies considered in Section~\ref{sec:mzsec}.
Fig.~\ref{fig:exhist} shows the trajectories of a few example galaxies in the
galaxy stellar mass - gas metallicity plane in the top panels, and the
corresponding evolution of the cold gas mass (green thick lines) and gaseous
metallicity (black lines) in the bottom panels. We have selected galaxies in a
narrow bin of galaxy stellar mass (${\rm log}\,{\rm M}_{\rm star}/{\rm
  M}_{\sun}=10.0-10.4$), and considered example galaxies whose trajectories in
the mass-metallicity plane are either relatively close (dashed lines) or
distant (dotted lines) from the present day best fit relation, for a significant
fraction of the galaxies lifetimes. The solid lines show instead intermediate
cases, and represent the most frequent behaviour we find for the model
  galaxies whose trajectories we have inspected by eye, independently of
galaxy stellar mass and offset from the best-fit relation. Our model galaxies
show a large variety of possible paths in the mass-metallicity plane, that are
determined by a complex interplay between different physical processes. While
the galaxy stellar mass increases monotonically with time, the gaseous
metallicity experiences significant oscillations. Only for a minority of the
galaxies inspected, we find evolutionary tracks that oscillate around the
mass-metallicity relation that is predicted at $z=0$. Most of the tracks lie
even below the mass-metallicity relation predicted at higher redshift (in fact,
the evolution is rather weak, as will be discussed in detail in a forthcoming
paper). The variety of tracks predicted by our model reflects the complex
interplay between the different physical processes driving the chemical
enrichment of the inter-stellar medium.

\begin{figure*}
\centering
\resizebox{16cm}{!}{\includegraphics{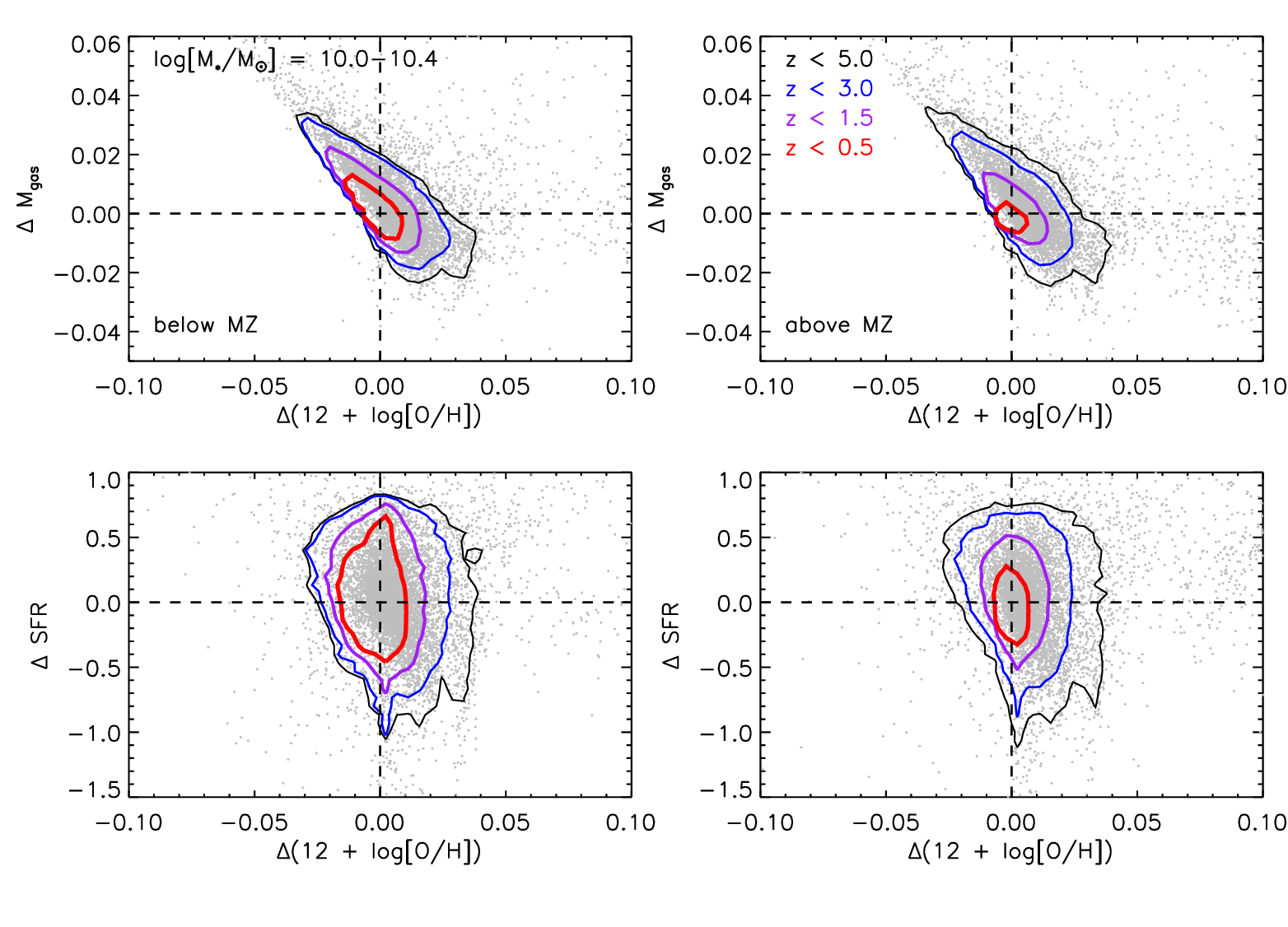}}
\caption{Percentage variations of cold gas mass (top panels) and star formation
  rate (bottom panels) as a function of variations of cold gas metallicity, for
  galaxies with stellar mass between $10^{10.2}$ and $10^{10.4}\,{\rm
    M}_{\odot}$. Left and right panels are for galaxies offset below and above
  the model best fit mass-metallicity relation (the lower/upper 15th
  percentile), respectively. Gray symbols represent a random subset of
    10,000 individual variations, covering the redshift range from zero to the
    highest redshift at which progenitors of model galaxies are found.
  Contours show regions including 90 per cent of the points at $z\,<\,5.0,\,
  3.0,\, 1.5,\, {\rm and}\, 0.5$, in order of increasing thickness and
  colour-coded as indicated in the legend. Dashed lines are plotted to guide
  the eye.
  \label{fig:dg}}
\end{figure*}

The bottom panels of Fig.~\ref{fig:exhist} show that metallicity variations are
strongly correlated with variations of the cold gas mass. Specifically,
negative variations of gas mass typically correspond to positive variations of
metallicity, i.e. the gaseous metallicity increases when the gas decreases. On
the other hand, positive variations of gas mass can lead to both an increase or
a decrease of the gaseous metallicity.

To evaluate statistical correlations between variations of different physical
quantities in our model, we consider all galaxies in the same narrow bin of
galaxy stellar mass (${\rm log}\,{\rm M}_{\rm star}/{\rm M}_{\sun}=10.0-10.4$),
and store all information about variations of cold gas mass, star formation
rate, and gaseous metallicity measured by following the main branch (i.e. the
branch traced by the most massive progenitor at each redshift).
Fig.~\ref{fig:dg} shows the percentage variations of cold gas mass (top panels)
and star formation rates (bottom panels) as a function of the gaseous
percentage metallicity variations, measured between two subsequent snapshots of
the model outputs. Gray symbols represent a random subset of 10,000 individual
variations, covering the redshift range from zero to that of the first main
progenitors of the model galaxy sample considered. Contours enclose the regions
containing 90 per cent of the percentage variations corresponding to $z<5.0,
3.0, 1.5, $ and $0.5$ (in order of increasing line thickness, and colour-coded
as indicated in the legend). We have splitted the sample according to their
offset from the best-fit mass-metallicity relation (left and right panels are
used for galaxies in the lower and upper 15th percentile of the distribution,
respectively). The figure shows a very weak correlation between variations of
the cold gas metallicity and of the star formation rate (bottom panels), while
there is a clear correlation between variations of the gas metallicity and of
the cold gas mass (top panels). The correlation becomes tighter with decreasing
redshift and for negative variations of the metallicity, and is such that
decreases/increases of cold gas lead on average to increases/decreases of the
cold gas metallicity. The bottom panels of Fig.~\ref{fig:dg} show that
increases of the star formation rate can lead both to positive and negative
variations of the cold gas metallicity. The latter case is expected e.g. during
galaxy mergers with gas-rich but low-mass satellites, that in our model would
trigger a starburst after the cold gas has been depleted due to the merger.
Finally, Fig.~\ref{fig:dg} also shows that the differences between galaxies
that are offset above or below the mass-metallicity relation become clearer, in
terms of the differences considered, with decreasing redshift: galaxies above
the mass-metallicity relation tend to have smaller percentage variations of gas
mass and star formation rates than their counter-parts below the relation. The
narrow tilted features that are visible in the lower-right regions of the
bottom panels, when looking at the highest redshift contour shown, correspond to
variations of the star formation rate following merger-driven bursts. In {\sc
  GAEA}, these are modelled using the `collisional starburst' prescription as
introduced by \citet*{Somerville_etal_2001}, where the mass of new stars formed
is a fixed fraction, that depends on the mass ratio of the merging galaxies, of
the gas mass available.

We have verified that the trends shown in Fig.~\ref{fig:dg} do not depend
significantly on galaxy stellar mass, but there is an increase of the relative
variations of cold gas mass and metallicity for decreasing galaxy mass.

\begin{figure}
\centering
\resizebox{8.7cm}{!}{\includegraphics{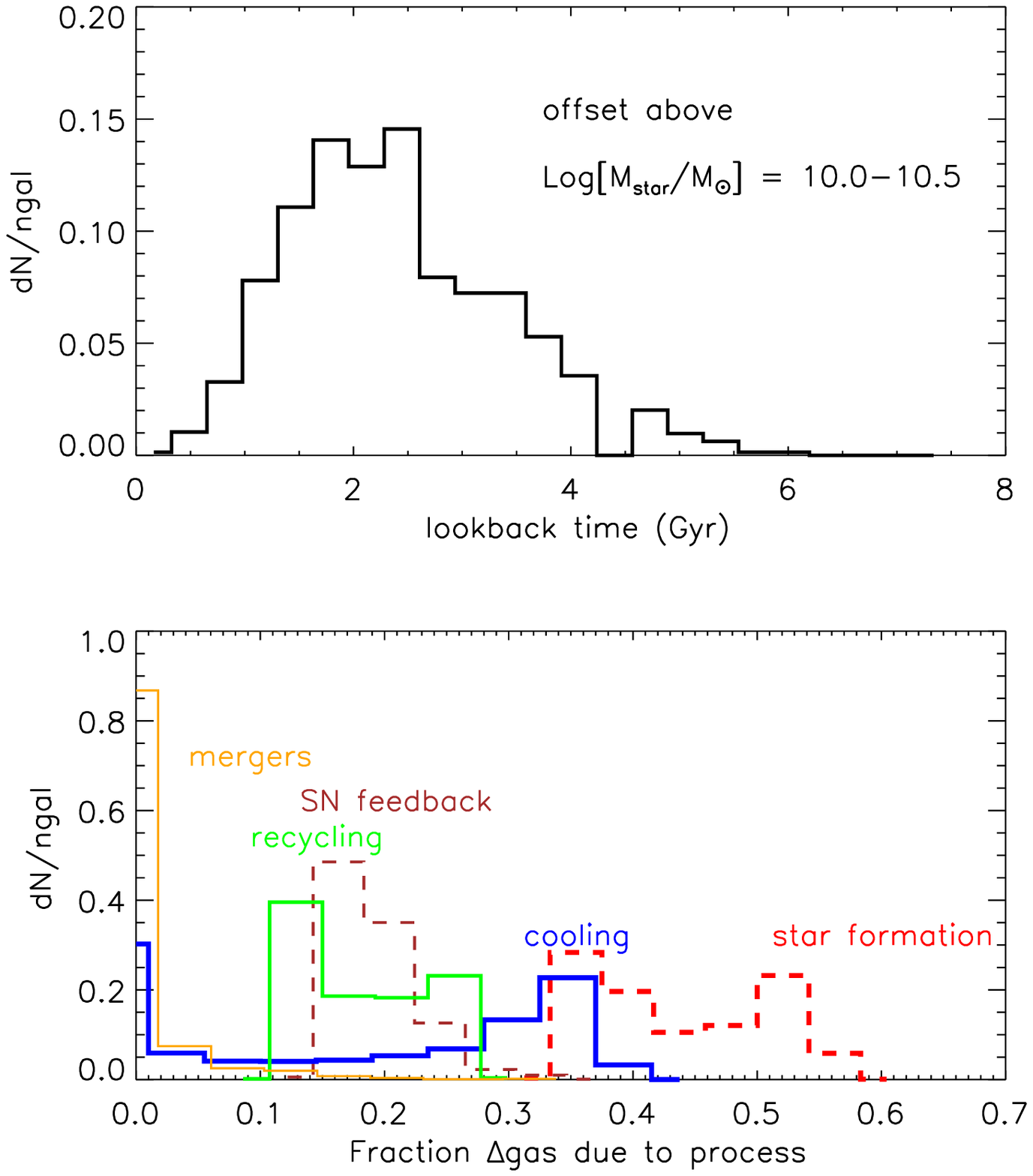}}
\caption{The top panel shows the distribution of the lookback times when the
  main progenitor crosses (for the first time going backward in time) the
  best-fit relation found for model galaxies at $z=0$. The bottom panel shows
  the fractional contributions of different physical processes to the absolute
  variation of cold gas in model galaxies, between the crossing time defined
  above and today. Contributions from processes that cause negative/positive
  variations of the cold gas mass are shown with dashed/solid lines. Increasing
  line thickness is used for stellar feedback, star formation (that cause
  negative variations of cold gas) and mergers, recycling and cooling (that lead
  to positive variations of cold gas).
  \label{fig:anabove}}
\end{figure}

Using the detailed information saved from our model, we can now address the
following question: which physical processes are determining positive and
negative variations of gas, and therefore offset model galaxies below or above
the mass-metallicity relation? To address this question, we have considered
again galaxies in a narrow mass bin, and only examined those that are at the
extreme of the distribution (above the 95th percentile, or below the 5th
percentile). This selection is aimed at having clearer differences between
the two populations, but we have verified that our results do not change
qualitatively when using a different selection. Below, we comment explicitly on
how the trends found change as a function of galaxy stellar mass. 

Let us start from galaxies that deviate above the mass metallicity relation.
For each model galaxy, we have followed back its main progenitor and recorded
the first time when the main progenitor crosses the best-fit relation of model
galaxies at $z=0$ (see top left panel of Fig.~\ref{fig:exhist}). We have also
recorded the variations of gas, between this time and $z=0$, due to: star
formation, stellar feedback, gas recycling, gas cooling, and galaxy mergers (in
our model, feedback from active galactic nuclei does not affect the amount of
cold gas in model galaxies, i.e. does not trigger galactic winds). Among the
processes considered, star formation and stellar feedback cause negative
variations of cold gas, while all other processes cause positive variations of
cold gas mass. In the top panels of Fig.~\ref{fig:anabove}, we show the
distribution of the lookback times when galaxies with stellar mass between
${\rm log}({\rm M}_{\rm star}/{\rm M}_{\odot})= 10$ and
${\rm log}({\rm M}_{\rm star}/{\rm M}_{\odot})=10.5$ (and above the 95th
percentiles of the metallicity distribution) cross the best-fit relation found
for model galaxies at $z=0$. The distribution has a clear and well defined peak
around $\sim 2$~Gyrs ago (i.e. $z\sim 0.17$ in our cosmology), with a tail at
earlier epochs.

We find that almost all galaxies selected using the above criteria ($\sim 97$
per cent of them) experience a negative variation of gas between the time they
cross the mass-metallicity relation and the present time. These vary between
$\sim 7.5\times 10^6\,{\rm M_{\odot}}$ and $\sim 8.5\times10^9\,{\rm
  M_{\odot}}$, with a median of $\sim 2.5\times 10^{9}\,{\rm M_{\odot}}$.  The
bottom panel of Fig.~\ref{fig:anabove} shows the fractional contributions to
the absolute variation of gas by different physical processes: the largest
contribution, with an average of $\sim 44$ per cent, comes from star formation.
For a very small fraction of the galaxies considered (including the $\sim 3$
per cent that experience positive variations of cold gas mass since crossing
time), gas cooling contributes slightly more than star formation ($\sim 38$
against $34$ per cent). Stellar feedback and gas recycling contribute to $\sim
18-19$ per cent of the absolute variation of cold gas mass (the former reduces
the cold gas in model galaxies, while the latter increases it).  Finally,
galaxy mergers represent a negligible contribution for virtually all galaxies
considered.

We have also tested if and how much results change as a function of galaxy
stellar mass considering a lower (${\rm log}({\rm M}_{\rm star}/{\rm
  M}_{\odot})= 9.5 - 10.0$) and higher (${\rm log}({\rm M}_{\rm star}/{\rm
  M}_{\odot})=10.5 - 11.0$) stellar mass bin. We find some weak trends as a
function of galaxy mass: for lower/higher mass galaxies, the tail towards
early crossing times of the mass-metallicity relation tends to be more/less
pronounced, respectively. The fractional contribution from star formation is
largest ($\sim 53$ per cent) for the most massive sample, where we find that
100 per cent of the galaxies considered experience a negative variation of gas
between the time when their progenitors cross the mass-metallicity relation and
today. The equivalent of Fig.~\ref{fig:anabove} for the other two galaxy mass
bins considered are provided in Appendix~\ref{sec:massdep}.

As discussed earlier, negative variations of gas are always associated with
positive variations of metallicity. In our model, stellar feedback removes a
fraction of the cold gas reheating or ejecting it. We assume that this occurs
leaving the metallicity of the cold gas component unchanged which means that,
in our model, positive variations of metallicity can only be caused by star
formation. Therefore, {\it late star formation, and the corresponding metal
  production, is primarily responsible for the offset of galaxies above the
  mass metallicity relation}.

\begin{figure}
\centering
\resizebox{8.7cm}{!}{\includegraphics{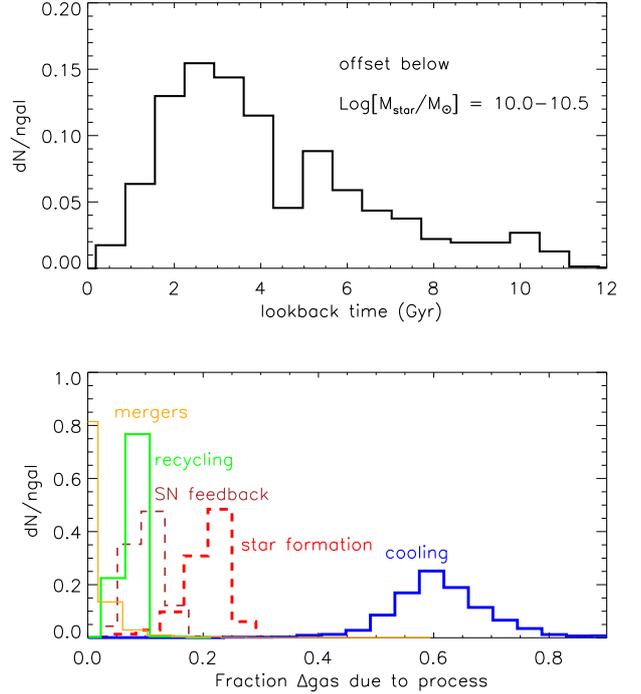}} 
\caption{The top panel shows the distribution of the lookback times
  corresponding to the most recent maximum value of the cold gas metallicity.
  The bottom panel shows, as in Fig.~\ref{fig:anabove}, the fractional
  contributions of different physical processes to the absolute variation of
  cold gas, between the time defined above and today. As for
  Fig.~\ref{fig:anabove}, we have highlighted with dashed histograms the
  contributions of physical processes causing negative variations of the cold
  gas.
  \label{fig:anbelow}}
\end{figure}

Let us now move to galaxies that are offset below the mass-metallicity relation.
We again consider galaxies with stellar mass between
${\rm log}({\rm M}_{\rm star}/{\rm M}_{\odot})= 10$ and
${\rm log}({\rm M}_{\rm star}/{\rm M}_{\odot})=10.5$, but now select only those
that are below the 5th percentile of the metallicity distribution. In this case,
we trace the main progenitor of each model galaxy backward in time, up to the
time corresponding to the most recent maximum of the cold gas metallicity. For
the example galaxies shown in Fig.~\ref{fig:exhist}, these times are marked as
filled circles. We then compute the variations of gas due to different
physical processes between the time just defined and present. Results are shown
in Fig.~\ref{fig:anbelow}. The distribution of the times corresponding to the
last maximum value of the cold gas metallicity peaks at $\sim 2.5$~Gyrs, but
has a broad tail that extends up to 10~Gyrs ago.

All galaxies considered in this case experience a positive variation of the
cold gas, with the largest contribution ($\sim 60$ per cent on average) coming
from gas cooling. The contributions from star formation, stellar feedback and
gas recycling are significantly lower (and also lower than the corresponding
contributions found for galaxies that are offset above the mass-metallicity
relation). Also in this case, galaxy mergers represent a negligible contribution
to the absolute variation of cold gas.

\begin{figure*}
\centering
\resizebox{18cm}{!}{\includegraphics{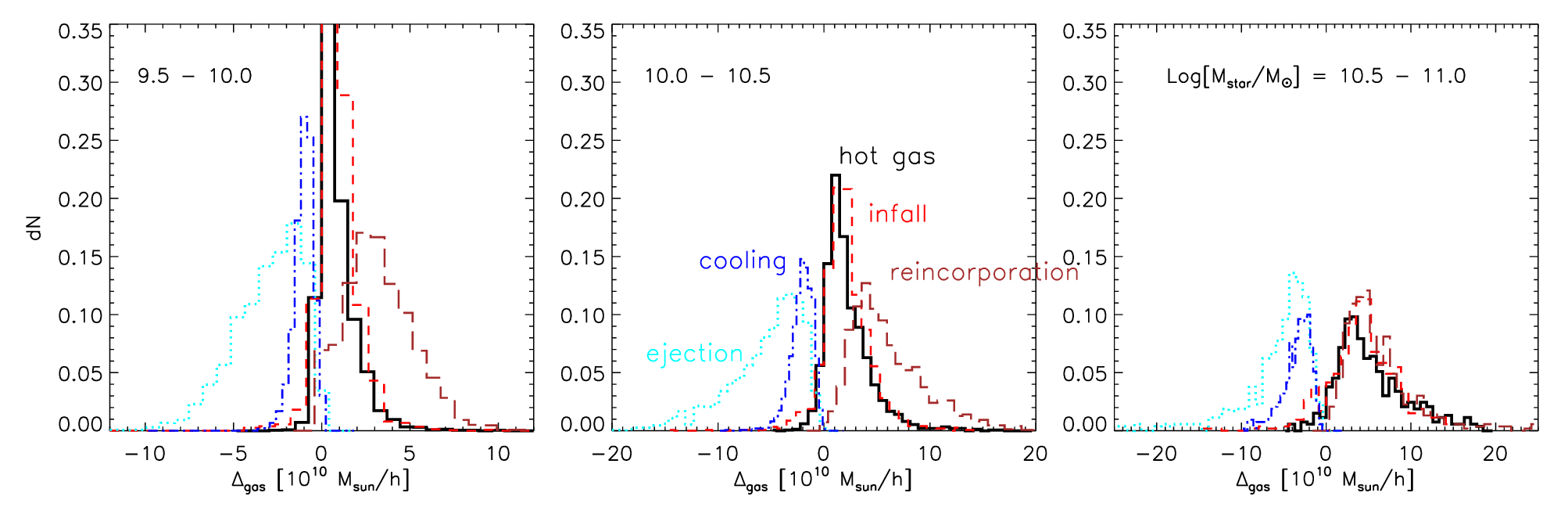}} 
\caption{Variations of hot gas (black solid histograms), cosmologically
  infalling gas (dashed red), ejected gas (dotted cyan), cooling gas
  (dot-dashed blue) and re-incorporated gas (long-dashed brown), between the
  time corresponding to the last maximum metallicity value and the present
  time. The three panels correspond to different galaxy stellar mass bins, as
  indicated in the legend.
\label{fig:ancool}}
\end{figure*}

We have verified if there is a dependence on galaxy stellar mass: for less and
more massive galaxies, the tail of the distribution of the times of maximum
metallicity becomes more and less pronounced at early cosmic epochs,
respectively. In addition, the contributions from star formation and recycling
tend to decrease/increase for less/more massive galaxies, against a
larger/smaller fractional contribution from gas cooling. For the least massive
bin considered, a very small fraction of the galaxies ($\sim 3$ per cent)
experience a negative variation (i.e. a reduction) of the cold gas available.
For these galaxies, the fractional contribution of star formation is actually
larger than that of gas cooling (i.e. they behave like the galaxies analysed in
Fig.~\ref{fig:anabove}). 

As discussed earlier, positive variations of cold gas can be associated with
both positive and negative variations of metallicity. Our analysis indicates
that, in most cases, {\it offsets below the mass metallicity relation can be
  explained by dilution of cold gas due to late gas cooling}.

It should be noted that the galaxy samples used for Fig.~\ref{fig:anabove} and
Fig.~\ref{fig:anbelow} do not have the same distribution of galaxy stellar mass.
In particular, galaxies offset above the mass-metallicity relation tend to be
on average more massive than galaxies offset below the relation. We have
verified that the trends discussed above remain qualitatively the same when
using mass matched samples. In this case, we find that galaxies that are offset
below the mass-metallicity relation at $z=0$ tend to reside in slightly more
massive haloes (${\rm log \, M_{halo}/M_{\sun}}\sim 11.85$) than galaxies of the
same mass that are found above the mass-metallicity relation (for these
galaxies, ${\rm log \, M_{halo}/M_{\sun}}\sim 11.6$). This halo mass difference
appears at low redshift ($\lesssim 0.4$) and leads to an increase of the gas
cooling (and therefore of the cold gas masses) in galaxies below the
mass-metallicity relation. This late gas cooling also leads to larger radii,
lower molecular to atomic mass ratios, and lower star formation rates.  We find
that these trends are independent of the stellar mass bin considered, but more
prominent for low and intermediate mass galaxies. This suggests that AGN
feedback, whose efficiency instead increases with increasing galaxy mass, plays
a negligible role in suppressing cooling for galaxies that are found above the
mass-metallicity relation. Therefore, the scatter above the relation is due to
a lack of late gas accretion, rather than to an excess of star formation.

One final question that we can address using results from our model is to what
extent late gas cooling is due to cosmological infall or re-incorporation of
gas that has been previously ejected due to stellar feedback. We show the
relevant information in Fig.~\ref{fig:ancool}. The three panels correspond to
three different stellar mass bins considered. In each panel, we show the
distributions of the variations of the hot gas (solid black histograms), the
ejected gas (dotted cyan), the cooling gas (dot-dashed blue), the infalling gas
(dashed red), and the re-incorporated gas (long-dashed brown). As for
Fig.~\ref{fig:anbelow}, variations are defined between the time corresponding
to the last maximum metallicity value and the present time. The figure shows
that for the lowest and intermediate stellar mass bins considered, the largest
contribution is coming from re-incorporation of gas that has been previously
ejected outside the galaxy haloes. For the most massive bins, the absolute
value of the contribution from cosmological infall increases, and the relative
importance of this channel becomes comparable to that of gas re-incorporation.
The behaviour seen in Fig.~\ref{fig:ancool} is a consequence of our assumption
that re-incorporation time-scales are inversely proportional to halo mass
\citep[][and references therein]{Hirschmann_etal_2016}. As discussed in that
work (see their Fig.~4), in our model, gas re-accretion is suppressed at early
cosmic epochs and is delayed to progressively lower redshifts for galaxies of
decreasing stellar mass. Since the ejection occurs at much earlier times than
re-incorporation, the material that is re-accreted is relatively metal-poor,
leading to a dilution of the cold gas metallicity. In particular, we find that
the average  metallicity of the re-incorporated gas is about $0.4-0.5$ times
that of the hot gas for re-incorporation events that occur below $z\sim 1$.

\section{Discussion and Conclusions}
\label{sec:discconcl}

In this paper, we have taken advantage of the state-of-the-art GAlaxy Evolution
and Assembly (GAEA) semi-analytic model \citep{Hirschmann_etal_2016} to analyse
the origin of secondary dependencies in the observed correlation between galaxy
stellar mass and cold gas metallicity. Our model reproduces well the observed
evolution of the relation up to $z\sim 2$
\citep{Hirschmann_etal_2016,Xie_etal_2017}, and includes an explicit treatment
for the partition of the cold gas in its atomic and molecular components
\citep{Xie_etal_2017}. This has allowed us to carry out a detailed comparison
with the trends observed in the local Universe, where the existence of a
secondary dependence in the mass-metallicity relation has been studied as a
function of different physical properties, including galaxy star formation rate
and gas content (both atomic and molecular).

Our model reproduces relatively well, both in terms of normalization and shape,
measurements of the mass-metallicity relation in the local Universe based on
strong emission lines. The model also reproduces qualitatively well the trends
measured as a function of star formation rate and of different gas-phase
components \citep{Bothwell_etal_2016,Brown_etal_2018,Curti_etal_2020}.
Specifically, the predicted offsets from the mass-metallicity relation depend
weakly on the star formation rate and mass of the molecular gas, while the
correlation is stronger with the atomic gas mass. 
Therefore, in agreement with what is inferred in the local Universe, the cold
gas content of our model galaxies (whose largest fraction is represented by the
atomic phase) can be considered as the third parameter governing the scatter of
the mass-metallicity relation. 

As argued in previous work \citep[see e.g. discussion in ][and references
  therein]{Brown_etal_2018}, the observed trends can be explained with 
fluctuations of the gas accretion rates: a decrease of the gas supply leads to
an increase of the gas metallicity as star formation progresses, while an
increase of the available gas (due to e.g. gas cooling following cosmological
infall, or galaxy-galaxy mergers) would generally lead to a dilution of the
cold gas metallicity. This picture is supported by several theoretical studies,
both based on the explicit {\it ansatz} of equilibrium between gas inflow,
  outflows, and star formation \citep[e.g.][]{Dave_etal_2012,Lilly_etal_2013},
and not assuming equilibrium \citep[e.g.][]{Dayal_etal_2013,Forbes_etal_2014}.
These studies, however, cannot explain the origin of the fluctuations of gas
supply, or distinguish between cosmological infall and re-accretion of gas that
has been previously ejected by galactic outflows.

In our model, galaxies that experience negative variations of gas due to star
formation, without the cold gas reservoir being replenished by additional gas
cooling, move along diagonal lines in a mass-metallicity plane (towards larger
masses and larger metallicities). As we have shown in the previous section, this
implies that galaxies that are offset above the mass-metallicity relation have
been driven there primarily by star formation. The contribution from stellar
feedback, that in our model only reduces the mass of cold gas but does not alter
its metallicity, is however not negligible. In a scenario where metals are
selectively lost via galactic winds, stellar feedback would reduce the cold gas
amount and also its metallicity. Thus, the contribution from star formation and
stellar feedback in determining variations of the cold gas metallicity would be
more similar, in absolute terms. The fractional contribution of star formation
to late changes of cold gas mass is stronger for more massive star-forming
galaxies, since they are characterized by higher star formation rates and
deeper potential wells that make gas removal by star formation less efficient.
The characteristic time-scales of the metallicity variation are, in this case,
relatively long because the star formation efficiency is generally only of the
order of few per cent. Therefore, galaxies that are found above the
mass-metallicity relation have `crossed' it relatively recently (for our model
galaxies, this happened on average $\sim 2$~Gyr ago). 

Positive variations of gas can lead to both positive and negative variations of
the cold gas metallicity. The former take place during early phases of galaxy
evolution and are driven by gas recycling and, to a minor extent, by mergers
with other gas-rich galaxies. The latter can occur in case of metallicity
dilution driven by either new supply of cold material (gas cooling) or by
mergers with lower mass, and therefore lower metallicity, galaxies. All our
model galaxies that are offset below the mass-metallicity relation in the local
Universe, have experienced positive variations of cold gas that are due to gas
cooling. More specifically, we find that the largest contribution to the newly
cooling gas is from material that has been previously ejected via stellar
feedback, for galaxies with stellar mass lower than $\sim 10^{10.5}\,{\rm
  M}_{\sun}$.  For more massive galaxies, the relative contribution
from this `ejected' component decreases and becomes comparable to that due to
cosmological infall of gas. This trend is, in our model, a consequence of
specific assumptions about the re-incorporation time-scales \citep[][and
  references therein]{Hirschmann_etal_2016}. Negative variations of the cold
gas metallicity occur on time-scales that are shorter than those associated
with positive variations of metallicity, and are essentially determined by the
cooling time-scales.

The stochasticity of halo merger histories leads to a wide range of possible
trajectories in the mass-metallicity plane. Albeit some general trends can be
found and an `average' behaviour can be described analytically, a full
understanding of the scatter in the observed mass-metallicity relation requires
cosmologically embedded galaxy formation models, like the one considered here.
In this work, we have focused on the local Universe and on the origin of
secondary dependencies in the mass-metallicity relation. In future work, we
plan to investigate if and how model predictions vary as a function of cosmic
time and how this evolution depends on the specific assumptions made for
stellar feedback.

\section*{Acknowledgements}
We thank Filippo Mannucci, Mirko Curti, and Toby Brown for providing us their
observational measurements in electronic format. FF acknowledges support from
the PRIN MIUR project ``Black Hole winds and the Baryon Life Cycle of Galaxies:
the stone-guest at the galaxy evolution supper'', contract 2017-PH3WAT. MH
acknowledges financial support from the Carlsberg Foundation via a ``Semper
Ardens'' grant (CF15-0384). We also thank Filippo Mannucci and Mirko Curti for
useful and constructive comments on a preliminary version of this manuscript.

\section*{Data availability}
The model data underlying this article will be shared on request to the
corresponding author.

An introduction to GAEA, a list of our recent work, as well as data files
containing published model predictions, can be found at
http://adlibitum.oats.inaf.it/delucia/GAEA/

\bibliographystyle{mnras}
\bibliography{mzgas} 

\appendix
\section{Dependence on galaxy stellar mass}
\label{sec:massdep}

\begin{figure}
\centering
\resizebox{8.7cm}{!}{\includegraphics{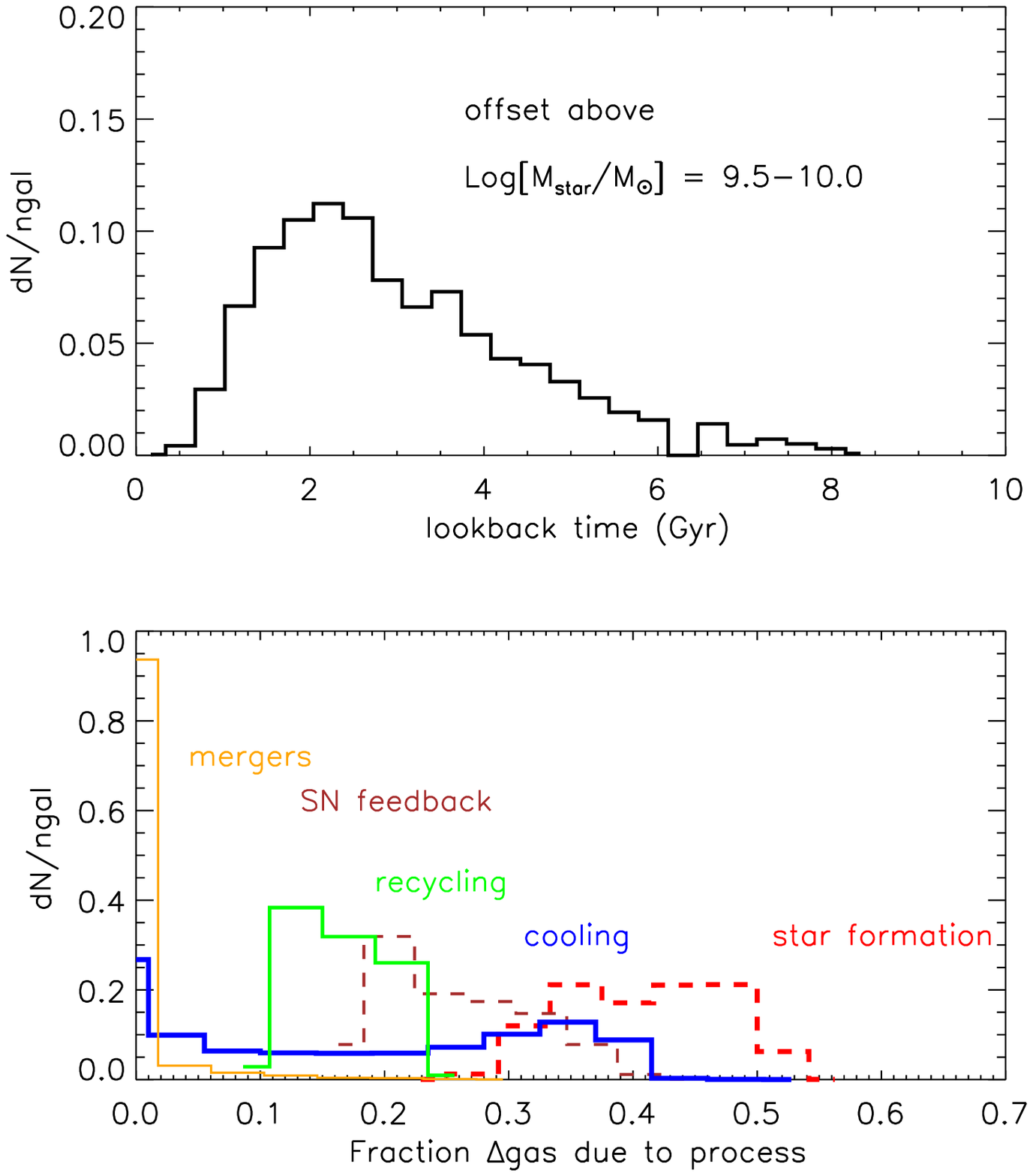}}
\caption{As in Fig.~\ref{fig:anabove}, but for galaxies with stellar mass
  between ${\rm log}({\rm M}_{\rm star}/{\rm M}_{\odot})= 9.5$ and ${\rm
    log}({\rm M}_{\rm star}/{\rm M}_{\odot})= 10$.
  \label{fig:anablm}}
\end{figure}

\begin{figure}
\centering
\resizebox{8.7cm}{!}{\includegraphics{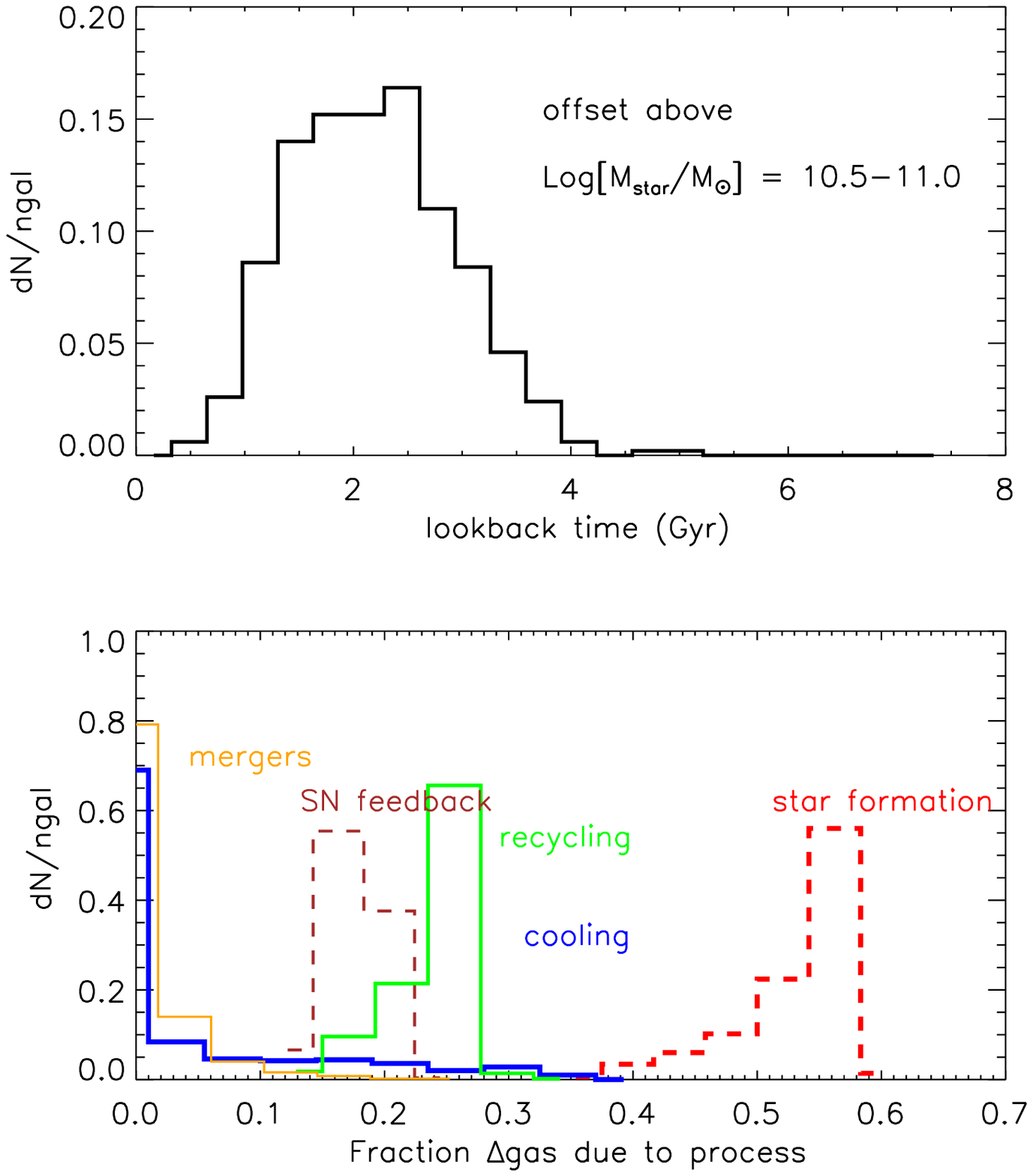}}
\caption{As in Fig.~\ref{fig:anabove}, but for galaxies with stellar mass
  between ${\rm log}({\rm M}_{\rm star}/{\rm M}_{\odot})= 10.5$ and ${\rm
    log}({\rm M}_{\rm star}/{\rm M}_{\odot})= 11$.
  \label{fig:anabhm}}
\end{figure}

In Section~\ref{sec:mz}, we have discussed how the trends discussed for
galaxies that are offset above or below the mass-metallicity relation depend on
galaxy stellar mass. For completeness, we provide in this appendix the
equivalents of Figs.~\ref{fig:anabove} and \ref{fig:anbelow} for galaxies more
and less massive than those considered in the main text. Specifically, the top
panels of Fig.~\ref{fig:anablm} and Fig.~\ref{fig:anabhm} show the distribution
of the lookback times when the main progenitors cross the best-fit
mass-metallicity relation, found for model galaxies at $z=0$. The bottom panels
show the fractional contributions of different physical processes to the
absolute variation of cold gas experienced by model galaxies between the
crossing time and today. Fig.~\ref{fig:anablm} corresponds to galaxies with
stellar mass between ${\rm log}({\rm M}_{\rm star}/{\rm M}_{\odot})= 9.5$ and
${\rm log}({\rm M}_{\rm star}/{\rm M}_{\odot})= 10$, while Fig.~\ref{fig:anabhm}
is for galaxies with stellar mass between
${\rm log}({\rm M}_{\rm star}/{\rm M}_{\odot})= 10.5$ and
${\rm log}({\rm M}_{\rm star}/{\rm M}_{\odot})= 11$.

\begin{figure}
\centering
\resizebox{8.7cm}{!}{\includegraphics{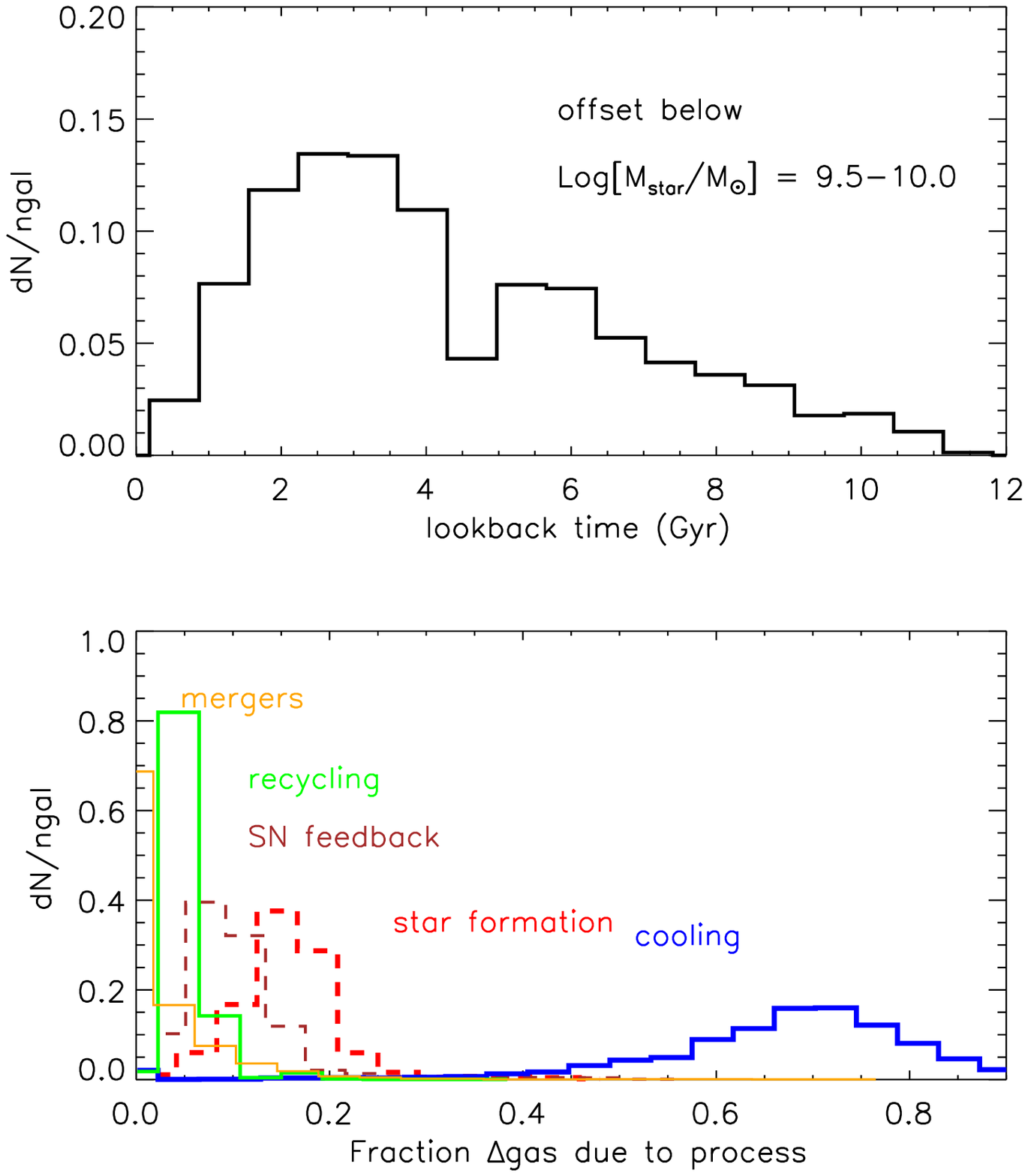}} 
\caption{As in Fig.~\ref{fig:anbelow}, but for galaxies with stellar mass
  between ${\rm log}({\rm M}_{\rm star}/{\rm M}_{\odot})= 9.5$ and ${\rm
    log}({\rm M}_{\rm star}/{\rm M}_{\odot})= 10$.
  \label{fig:anbelm}}
\end{figure}

\begin{figure}
\centering
\resizebox{8.7cm}{!}{\includegraphics{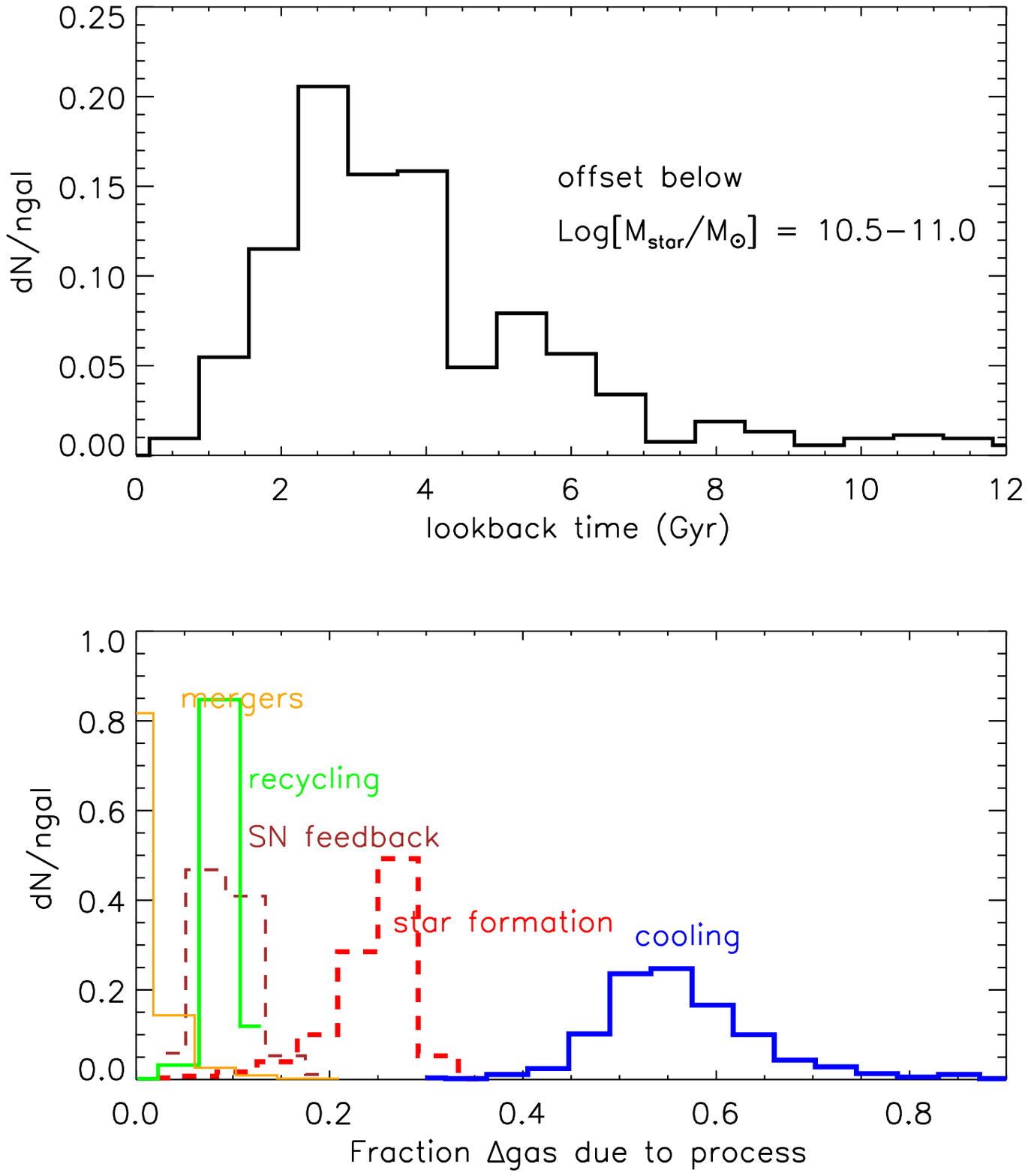}} 
\caption{As in Fig.~\ref{fig:anbelow}, but for galaxies with stellar mass
  between ${\rm log}({\rm M}_{\rm star}/{\rm M}_{\odot})= 10.5$ and ${\rm
    log}({\rm M}_{\rm star}/{\rm M}_{\odot})= 11$.
  \label{fig:anbehm}}
\end{figure}

Figs.~\ref{fig:anbelm} and \ref{fig:anbehm} show, for the same galaxy mass
bins, the distributions of the lookback times corresponding to the most recent
maximum values of the cold gas metallicity (top panels), and the fractional
contributions of different physical processes to the absolute variation of cold
gas experienced by model galaxies between this time and today (bottom panels).

\bsp	
\label{lastpage}
\end{document}